\documentclass[showpacs,amsmath,amsfonts,amssymb,aps,superscriptaddress]{revtex4}

\usepackage{makecell, tabularx}

\usepackage{booktabs}
\usepackage[dvipsnames, x11names]{xcolor}

\usepackage{comment}
\usepackage{enumitem}

\usepackage{graphicx}% Include figure files
\graphicspath{{figs/}}

\usepackage{float}
\usepackage{dcolumn}% Align table columns on the decimal point
\usepackage{bm}% bold math
\usepackage{mathrsfs}
\usepackage{amsmath}

\usepackage{amsfonts}
\usepackage{dsfont}
\makeatletter
\DeclareFontShape{OT1}{cmr}{m}{scit}{<->ssub * cmr/m/sc}{}
\makeatother

\begin{document}

\title{Noncommutative geometry–inspired wormholes supported by quasi-de Sitter and Chaplygin-like equations of state}

\author{Davide Batic}
\email{davide.batic@ku.ac.ae}
\affiliation{Department of Mathematics, Khalifa University of Science and Technology, PO Box 127788, Abu Dhabi, United Arab Emirates}

\author{Denys Dutykh}
\email{denys.dutykh@ku.ac.ae}
\affiliation{Department of Mathematics, Khalifa University of Science and Technology, PO Box 127788, Abu Dhabi, United Arab Emirates}

\author{Mark Essa Sukaiti}
\email{100064482@ku.ac.ae}
\affiliation{Department of Mathematics, Khalifa University of Science and Technology, PO Box 127788, Abu Dhabi, United Arab Emirates}
\date{\today}

\begin{abstract}
We construct static, spherically symmetric wormhole solutions with a nontrivial redshift function, inspired by noncommutative geometry, in which point sources are replaced by Gaussian smearing of minimal length, yielding a regular shape function. Within this framework, we derive model-independent relations that isolate the role of the redshift function in controlling the stress–energy components and the violation of the null energy condition (NEC). Negative or suitably tuned redshifts confine the exotic matter to a thin neighbourhood of the throat. We then reformulate this redshift engineering in matter terms through a quasi–de Sitter equation of state (EOS) with localised Gaussian or Lorentzian perturbations, obtaining minimally exotic wormholes that are regular, horizon-free, and asymptotically flat. Finally, we extend the analysis to a Chaplygin-like EOS, introducing a nonlinear coupling between pressure and density that yields redshift wells with possible local blueshift regions and tunable anisotropies governed by a certain nonlinearity parameter. Together, these results provide a unified and physically transparent framework for constructing traversable noncommutative-geometry–inspired wormholes with controlled, spatially localised exotic matter content.
\end{abstract}
\pacs{04.62.+v,04.70.-s,04.70.Bw} % PACS, the Physics and Astronomy
                              % Classification Scheme.
%\keywords{Suggested keywords}% Use the 'showkeys' class option if the keyword
                              %display desired
\maketitle

\section{Introduction}

Traversable wormholes, first popularised by Morris and Thorne \cite{Morris1988AJP}, are hypothetical bridges connecting distinct regions of spacetime. These solutions to Einstein’s field equations require stress-energy sources with exotic matter that violate the NEC to maintain an open throat and ensure traversability. The quest for physically reasonable models of such spacetimes has led to diverse approaches aimed at minimising or localising these violations.

One prominent direction involves the use of phantom fields, that is, scalar fields with reversed kinetic terms, which naturally produce negative energy densities near the throat and can thus support wormhole geometries \cite{Sushkov2005PRD, Lobo2005PRD, Lobo2005PRDa}. Although such models are often criticised for their quantum instabilities, which manifest as unbounded Hamiltonians, vacuum decay, or gradient instabilities \cite{Armendariz2001PRD, Caldwell2002PLB, Carroll2003PRD, Faraoni2005CQG}, they remain valuable as toy models for probing the structure and stability of wormholes. These investigations are typically conducted within the framework of semiclassical gravity, wherein the gravitational field is treated classically while quantum fields propagate on a fixed background. In this setting, one examines whether the renormalised expectation value of the quantum stress-energy tensor, $\langle \hat{T}_{\mu\nu} \rangle$, can provide the exotic energy densities required to sustain a traversable throat. The seminal work of \cite{Morris1988PRL, Hochberg1997PRD} demonstrated that vacuum fluctuations such as those of conformally coupled scalar fields can yield localised NEC violations sufficient to stabilise wormhole geometries. Alternative proposals have employed nonlinear electrodynamics as the matter source, where certain choices of Lagrangian can produce stress-energy tensors that violate the NEC in controlled regions \cite{Thibeault2006GRG, Eiroa2009PRD, Sharif2012JPSJ, Bronnikov1973APP, Bronnikov2018IJMPD, Shaikh2016PRD}. These models are particularly appealing because of their self-consistent coupling to general relativity and their origin in quantum field theory. Last but not least, within GR and semiclassical gravity, there is a complementary line of work aimed at quantifying and minimising the required amount of exotic matter, and at confining the NEC-violating region as close as possible to the throat (see \cite{Azreg2015JCAP, Fewster2005PRD, Zaslvskii2007PRD, Visser2003PRL}). The present work differs from these earlier constructions in that we work with a noncommutative-geometry-inspired smeared matter source and derive model-independent relations that promote the redshift function from an ansatz to a quantitative control parameter for the width and magnitude of the NEC-violating layer.

A different class of constructions exploits conformal symmetry or scale invariance, leading to solutions in which the effective stress-energy tensor violates the NEC only within a bounded domain near the throat \cite{Barcelo1999PLB, Mazharimousavi2009PLB}.  This includes so-called thin-shell wormholes, where two spacetimes are joined across a timelike hypersurface, and the exotic matter is confined to an infinitesimally thin shell \cite{Poisson1995PRD, Lobo2004CQG}. Such models make the NEC violation manifestly minimal in a distributional sense and have been widely used to probe stability under perturbations. Furthermore, quantum corrections via the renormalized stress-energy tensor in semiclassical gravity have shown that vacuum polarization effects near strong gravitational sources can generate NEC-violating components, potentially stabilizing small wormhole-like structures without invoking exotic matter in the form of phantom scalar fields, anisotropic fluids with pressure profiles engineered to produce NEC violation near the wormhole throat, non-minimally coupled scalar fields, or specific configurations dictated by nonlinear electrodynamics \cite{Ellis1973JMP, Ellis1974JMP, Hochberg1997PRL, Visser1990PLB}. In parallel, within modified‑gravity settings, one can combine noncommutative smearing with $f(R)$ dynamics to obtain wormholes that satisfy energy conditions in certain regimes, illustrating an alternative mechanism distinct from our present approach \cite{Baruah2023NA}. Overall, these frameworks offer critical insights into the structure of exotic matter and provide constraints that guide the development of more physically consistent wormhole models.

One particularly promising direction is represented by the introduction of noncommutative geometry, where the spacetime manifold is endowed with a fundamental discretisation scale \cite{Smailagic2003JPA, Smailagic2003JPAa, Nicolini2006PLB, Nicolini2010CQG}. In this framework, point-like structures are replaced by smeared distributions, effectively regularising curvature singularities. It is worth noting that the Gaussian-smeared energy density derived from noncommutative geometry yields black hole solutions with nonsingular cores and modified thermodynamic behaviour \cite{Nicolini2009IJMP, Casadio2008JHEP}. Inspired by this promising avenue, \cite{Garattini2009PLB} extended the noncommutative approach to construct wormhole geometries supported by a Gaussian source, demonstrating that the noncommutative geometry approach can reduce the severity of the energy condition violations typically required to support a traversable wormhole. This softening manifests in several ways: the NEC is violated more mildly, in the sense that quantities like $\rho+p_r$ become only slightly negative; the violation is confined to a narrow region around the throat, beyond which the spacetime satisfies the classical energy conditions; and the source of exoticity arises from a Gaussian-smeared matter distribution rooted in the geometry itself, rather than being imposed in an ad hoc or artificial manner. In the standard treatment, the redshift function, which encodes the lapse of proper time, is often taken to be constant. However, its detailed structure plays a crucial role in the geometry and stress-energy distribution of the wormhole. Motivated by this, recent work has explored nontrivial redshift profiles as a means of better controlling energy-condition violations \cite{Lobo2005PRD, Ovgun2019PRD}.

In contrast to modified-gravity realisations such as $f(R)$ models with noncommutative-inspired sources \cite{Baruah2023NA, Azreg2015JCAP}, our analysis remains within general relativity, where we show that the redshift function  $\Phi(r)$ can be tuned to confine the region of NEC violation.  Beyond revisiting the original noncommutative-inspired wormhole of \cite{Garattini2009PLB}, we establish model-independent relations that elevate the redshift function from an arbitrary ansatz to a quantitative control parameter for localising and minimising NEC violation.  In particular,
\begin{enumerate}
\item 
We isolate the redshift-independent NEC at the throat and derive a bound on $d\Phi/dr$ that governs NEC restoration away from it;
\item 
We obtain near-throat formulas connecting the slope and extent of the NEC-violating layer, together with an existence condition for a local crossing; and
\item 
We recast the geometry in matter terms through a quasi–de Sitter EOS with localised Gaussian or Lorentzian perturbations, leading to a master relation for $\Phi$ that generates horizon-free, asymptotically flat solutions with confined exoticity.
\end{enumerate}
To probe the robustness of this mechanism, we further introduce a Chaplygin-like EOS that adds a nonlinear coupling between pressure and density, inspired by the generalised Chaplygin gas. This extension enables us to investigate how nonlinear matter effects modify the redshift potential and anisotropies while preserving regularity and asymptotic flatness. The redshift families and matter models analysed here thus serve as complementary case studies validating the general framework and its predictive trends. The resulting construction provides a flexible, physically transparent framework for constructing traversable wormholes within noncommutative geometry.

The paper is organised as follows. In Section II, we review the Einstein field equations and the noncommutative geometry–inspired wormhole model, introduce the family of redshift functions, and analyse the corresponding violations of the NEC. Section III presents the matter-based formulation of the framework. More precisely, we derive and study the quasi–de Sitter EOS with localised Gaussian and Lorentzian perturbations, leading to minimally exotic, asymptotically flat solutions. Furthermore, we extend the analysis to a Chaplygin-like EOS, highlighting the effects of the nonlinear coupling between pressure and density on the redshift potential and anisotropies. Finally, Section IV summarises our findings and outlines possible directions for future work.

\section{THE NONCOMMUTATIVE GEOMETRY INSPIRED WORMHOLE}\label{Sec2}

\begin{figure}
  \includegraphics[scale=0.35]{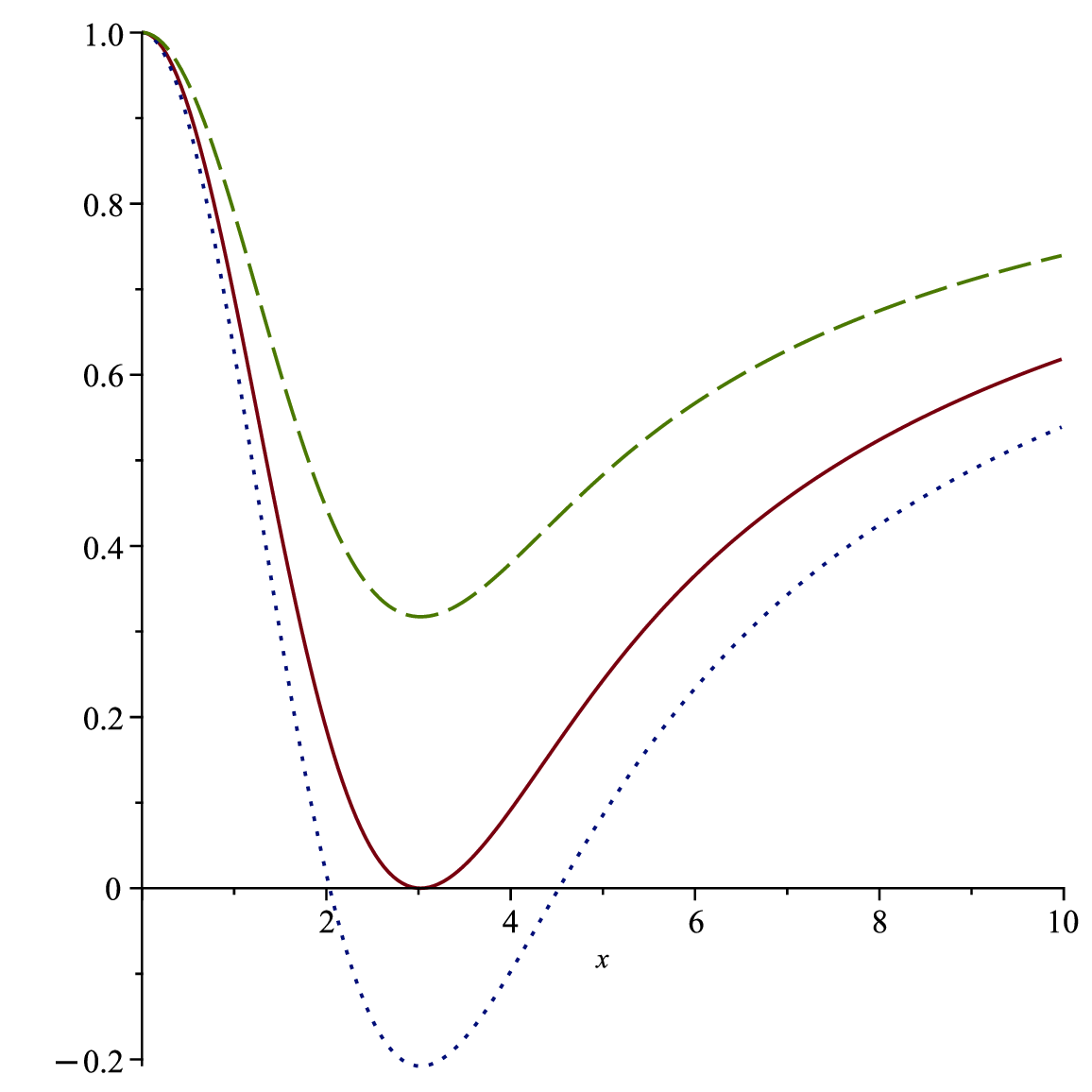}
  \caption{\label{fig0} Plot of the metric coefficient $g^{xx} = f(x)$, as defined in \eqref{fx}. The extremal case occurs at $\mu = \mu_e = 1.9041\ldots$ (solid line), where two coinciding throats are located at $x_e =3.0224\ldots$. For $\mu > \mu_e$, a non-extremal wormhole with two distinct throats is present (dotted line shown for $\mu = 2.3$). For $\mu < \mu_e$, the absence of throats indicates that the gravitational object is not a wormhole (dashed line shown for $\mu = 1.3$).}
\end{figure}

The spacetime of a noncommutative geometry-inspired wormhole is described by a static metric. Spherically symmetric metric, expressed in natural units where $c = G_N = 1$, and given by the line element \cite{Garattini2009PLB, Nicolini2010CQG}
\begin{equation}\label{metric}
  ds^2 = -e^{2\Phi(r)}dt^2+\frac{dr^2}{1-\frac{b(r)}{r}} + r^2d\vartheta^2 + r^2\sin^2{\vartheta}d\varphi^2, \quad \vartheta\in[0,\pi], \quad \varphi\in[0,2\pi).
\end{equation}
Here, $\Phi$ and $b$ are the redshift and shape functions, respectively, as defined in \cite{Garattini2009PLB, Nicolini2010CQG}. We use the standard areal–radius chart with domain $r\in[r_0,\infty)$. The full wormhole is obtained by glueing two copies of this exterior region at the throat $r=r_0$. Moreover, we introduce a timelike unit vector $u^{\alpha}$ (fluid 4‑velocity) and a radial spacelike unit vector $\ell^{\alpha}$, orthogonal to $u^{\alpha}$, satisfying 
\begin{equation}
u^{\alpha}u_{\alpha}=-1,\qquad \ell^{\alpha}\ell_{\alpha}=+1,\qquad
u^{\alpha}\ell_{\alpha}=0.
\label{eq:unit-orthogonality}
\end{equation}
In terms of these vectors, an anisotropic stress–energy tensor can be written as
\begin{equation}\label{emt}
   T^\alpha{}_\beta=(\rho+p_t)u^\alpha u_\beta+p_t\delta^\alpha{}_\beta+(p_r-p_t)\ell^\alpha\ell_\beta.  
\end{equation}
A convenient choice consistent with \eqref{eq:unit-orthogonality} is
\begin{equation}\label{utnorot}
  u^\alpha=e^{-\Phi(r)}\delta^\alpha{}_t, \quad \ell^\alpha=\sqrt{1-\frac{b(r)}{r}}\delta^\alpha{}_r.
\end{equation}
Hence, the mixed energy-momentum tensor can be represented in terms of the diagonal matrix
\begin{equation}
  T^\alpha{}_\beta=\text{diag}(-\rho(r),p_r(r),p_t(r),p_t(r)),
\end{equation}
where $\rho$ is the energy density, $p_r$ the radial pressure, and $p_t$ the tangential pressure measured orthogonally to the radial direction. By applying the Einstein field equations $G^\alpha{}_\beta  = 8\pi T^\alpha{}_\beta$ alongside the conservation equation $\nabla_\beta T^{\alpha\beta} = 0$, we obtain the following system of equations, where an  overdot denotes differentiation with respect to the radial coordinate
\begin{eqnarray}
  &&\dot{b}-8\pi r^2\rho=0,\label{eq1}\\
  &&2r(r-b)\dot{\Phi}-b-8\pi r^3 p_r=0,\label{eq2}\\
  &&r^2(r-b)\ddot{\Phi}+\left[r(r-b)\dot{\Phi}+\frac{b-r\dot{b}}{2}\right](1+r\dot{\Phi})-8\pi r^3 p_t=0,\label{eq3}\\
  &&r(\rho+p_r)\dot{\Phi}+r\dot{p}_r+2(p_r-p_t)=0.\label{eq4}
\end{eqnarray}
Using equations \eqref{eq1}--\eqref{eq3}, we can express the energy density, the radial and tangential pressures in terms of the redshift and shape functions as follows
\begin{eqnarray}
\rho&=&\frac{\dot{b}}{8\pi r^2},\label{ed}\\
p_r&=&\frac{1}{8\pi}\left[\frac{2}{r}\left(1-\frac{b}{r}\right)\dot{\Phi}-\frac{b}{r^3}\right],\label{pr}\\
p_t&=&\frac{1}{8\pi}\left(1-\frac{b}{r}\right)\left[\ddot{\Phi}+\dot{\Phi}^2-\frac{r\dot{b}-b}{2r(r-b)}\dot{\Phi}+\frac{\dot{\Phi}}{r}-\frac{r\dot{b}-b}{2r^2(r-b)}\right].\label{pt}
\end{eqnarray}
It is straightforward to verify that substituting equations \eqref{ed}-\eqref{pt} into the conservation equation \eqref{eq4} shows that it is identically satisfied, as expected from the consistency of the Einstein field equations. Notice that imposing a constant redshift implies that $\dot\Phi=0$ and the Einstein equations algebraically constrain the anisotropy. In particular, from \eqref{ed}–\eqref{pt} one finds
\begin{equation}\label{pteos}
p_t=-\frac{\rho+p_r}{2}.
\end{equation}
This closure relation is implicit in the constant redshift construction of \cite{Garattini2009PLB}, but was not stated explicitly there. As a corollary, at the throat, one has
\begin{equation}
p_t(r_0)=\frac{1-\dot{b}(r_0)}{16\pi r_0^2}>0,
\end{equation}
since $\dot{b}(r_0)<1$ by the flare–out condition. Finally, the noncommutative smearing makes all curvature invariants finite at the throat and throughout the domain (see \cite{Garattini2009PLB} for explicit forms).  In the framework of noncommutative geometry \cite{Smailagic2003JPA, Nicolini2006PLB, Nicolini2009IJMP}, the traditional notion of point-like sources is replaced by a smeared mass distribution governed by a Gaussian profile. This approach reflects the underlying noncommutative structure of spacetime and modifies the energy density as follows
\begin{equation}\label{density}
  \rho(r)=\frac{M}{(4\pi\theta)^{3/2}}e^{-\frac{r^2}{4\theta}},
\end{equation}
where $M$ represents the total mass and $\theta$ is a positive parameter encoding the effects of spacetime noncommutativity. As emphasised in \cite{Nicolini2009IJMP}, the noncommutative scale is constrained to $\sqrt{\theta}< 10^{-16}$ cm, implying that the Gaussian profile exhibits a sharply peaked distribution. Consequently, the mass remains highly localised, although the delta-function singularity is effectively regularised. Throughout this work, we adopt the energy density given by \eqref{density}. Inserting this into \eqref{ed}, the shape function $b$ can be directly integrated, following the approach in \cite{Garattini2009PLB, Batic2025CQG}, yielding
\begin{equation}\label{shapefcn}
  b(r)=\frac{4M}{\sqrt{\pi}}\gamma\left(\frac{3}{2}, \frac{r^2}{4\theta}\right), \quad
  \gamma\left(\frac{3}{2}, \frac{r^2}{4\theta}\right) = \int_0^{r^2/4\theta} dt \, \sqrt{t} e^{-t},
\end{equation}
where $\gamma$ is the lower incomplete Gamma function. Moreover, we restrict our analysis to the domain $r \geq r_0$, where $r_0$ denotes the throat radius, and employ the shape function defined in \eqref{shapefcn} in conjunction with the redshift functions presented in Table~\ref{table:Phi}. In the following, it is convenient to introduce the function  
\begin{equation}\label{f}  
  f(r) =1-\frac{b}{r}= 1 - \frac{4M}{\sqrt{\pi}r}\gamma\left(\frac{3}{2}, \frac{r^2}{4\theta}\right),  
\end{equation}  
as this representation, resembling the $g_{00}$ metric coefficient of a noncommutative geometry-inspired Schwarzschild black hole \cite{Nicolini2006PLB}, simplifies the analysis of the throat location. More precisely, the procedure can be carried out in a manner analogous to the study of the event and Cauchy horizons in the aforementioned black hole \cite{Nicolini2006PLB, Batic2024EPJC}. In the following, we work with the dimensionless variables $(x,\mu)$, where 
\begin{equation}\label{newv}
  x = \frac{r}{\sqrt{\theta}}, \quad \mu = \frac{M}{\sqrt{\theta}}.
\end{equation}
If we use the identities \cite{Abramowitz1972}
\begin{equation}\label{ids}
  \gamma\left(\frac{3}{2}, w^2\right) = \frac{1}{2} \gamma\left(\frac{1}{2}, w^2\right) - w e^{-w^2}, \quad \gamma\left(\frac{1}{2}, w^2\right) = \sqrt{\pi} \, \mathrm{erf}(w),
\end{equation}
we can rewrite $f$ in \eqref{f} as
\begin{equation}\label{fx}
  f(x) =1-\frac{b}{x}= 1 -\frac{2\mu}{x}\mathrm{erf}\left(\frac{x}{2}\right)+\frac{2\mu}{\sqrt{\pi}} e^{-x^2/4},
\end{equation}
where \(\mathrm{erf}\) denotes the error function. The key insights derived from Figure~\ref{fig0}, which depicts $f$ as a function of the rescaled radial coordinate $x$, are summarized as follows
\begin{itemize}
  \item \underline{Case 1} ($\mu < \mu_e = 1.9041\ldots$): No real roots of $f$ exist in this regime, indicating the absence of a wormhole throat.  
  \item \underline{Case 2} ($\mu = \mu_e$): A degenerate configuration arises in which two coincident zeroes appear at $x_e =3.0224\ldots$.  
  \item \underline{Case 3} ($\mu > \mu_e$): Two distinct throats emerge at $x_\pm$. We identify the physical throat with the root $x_0 = x_+$, since it increases monotonically with the mass parameter $\mu$, while $x_{-}\to 0$. This choice is consistent with the physical expectation that the throat radius should increase with the mass parameter.
\end{itemize}
It is important to note that only the non-extremal configurations analysed in this study satisfy the flare-out condition at the throat $b^{'}(x_0)<1$ \cite{Morris1988AJP}, thereby qualifying as traversable wormholes. Here, a prime denotes differentiation with respect to $x$. In contrast, the extremal case $\mu = \mu_e$ corresponds to a degenerate scenario in which the two throats coalesce into a single throat at $x_e$. This configuration saturates the flare–out inequality, with $b'(x_e) = 1$ (see Table~I), and therefore does not represent a traversable wormhole in the Morris–Thorne sense. From the global point of view, it plays the role of a marginal, non-traversable limit separating the regime without throats ($\mu<\mu_e$) from the
family of genuine wormholes ($\mu>\mu_e$). In what follows, we use this extremal geometry
only as a reference point and concentrate on non-extremal configurations with $\mu>\mu_e$. In this regime, the absolute mass scale $M$ is determined by the minimal length $\sqrt{\theta}$ rather than by astrophysical scales. Concretely, we consider configurations with masses just above extremality, $M\simeq \mu_e\sqrt{\theta}$. This makes it clear that our models are noncommutative-geometry–inspired and intrinsically microscopic. Nowhere do we assume astrophysical masses of order $M\simeq M_\odot$, where $M_\odot$ denotes the solar mass. As the mass increases, however, the solutions gradually approach their classical counterparts. As shown in Table~\ref{table:throat}, for $M\sim 10^3\sqrt{\theta}$ the throat position tends to the Schwarzschild radius, signalling the onset of classical behaviour.
\begin{table}[ht]
%\centering
\caption{Representative numerical values of the throat location $x_0$ and the corresponding evaluation of the flare–out condition for several values of the rescaled mass parameter $\mu$. The table lists $b'(x_0)$, $b''(x_0)$, and the quantity $b'(x_0)-1$, which measures the deviation from the flare–out bound $b'(x_0)<1$.}
\label{table:throat}
\vspace*{1em}
\begin{tabular}{||c|c|c|c|c||}
\hline\hline
$\mu$              & $x_0$           & $b^{'}(x_0)$    &$b^{''}(x_0)$   &   $b^{'}(x_0)-1$\\ [0.5ex]
\hline\hline
$\mu_e$            & $x_e$           & $1$   & $-0.8495$ &   $0$\\
$1.905$            & $3.0803$   & $0.9513$   & $-0.8475$ &   $-0.0487$\\
$1.950$            & $3.4659$   & $0.6559$   & $-0.7582$ &   $-0.3441$\\
$2.5$              & $4.9685$   & $0.0727$   & $-0.1513$ &   $-0.9273$\\
$10^3$             & $2\cdot 10^3$   & $0$             & $0$            &   $-1$\\[1ex]
\hline\hline 
\end{tabular}
\end{table}

\begin{figure}
  \includegraphics[scale=0.35]{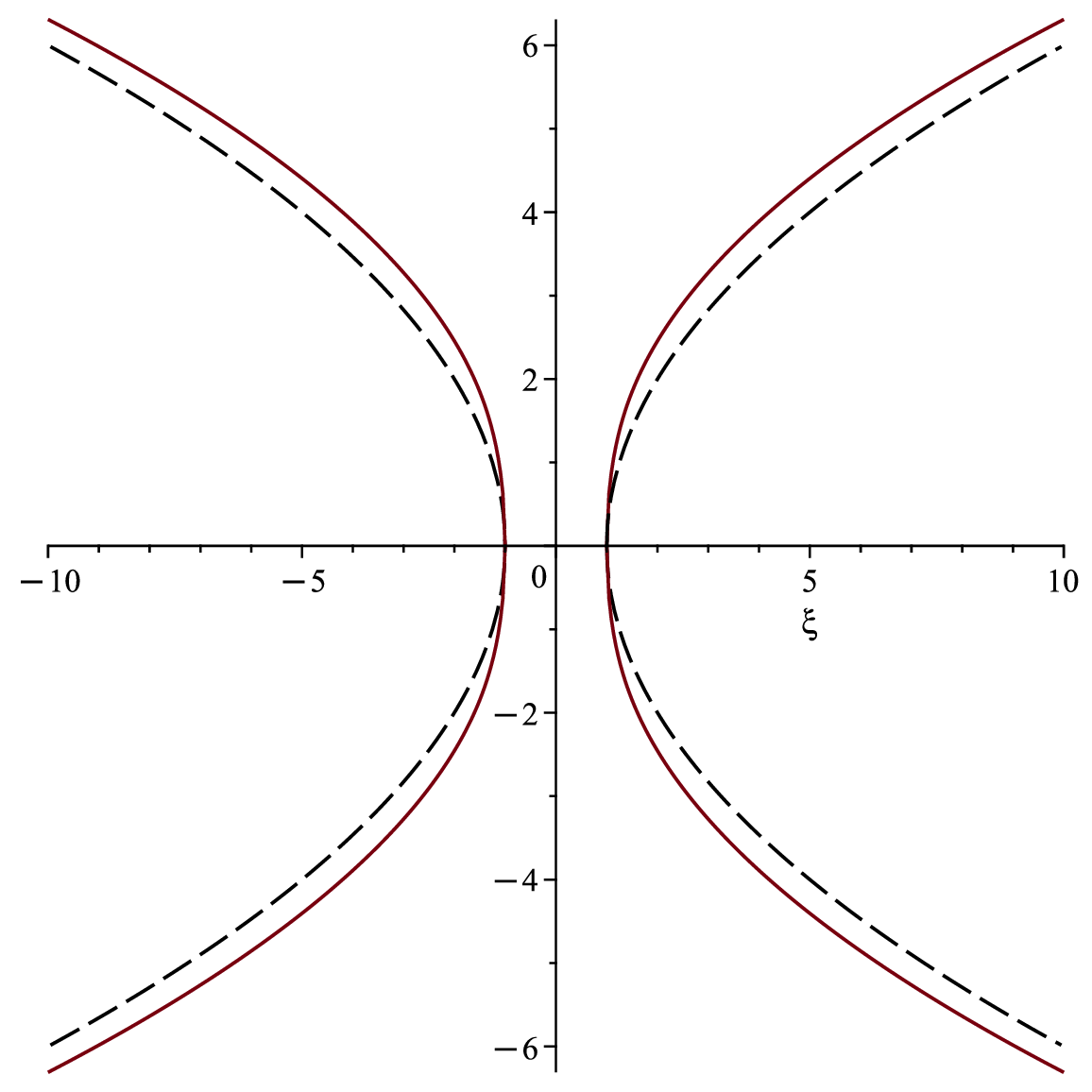}
  \caption{\label{profileb} Embedding diagram for the noncommutative wormhole with $\mu = 1.95$, shown in comparison to the corresponding Morris--Thorne wormhole (dashed line), as viewed in profile. To obtain the full three-dimensional geometry, the diagram must be revolved around the vertical $\widetilde{z}$-axis.}
\end{figure}

Furthermore, by introducing the rescaled variables $\xi = x / x_0$ and $\widetilde{z} = z / x_0$, and fixing $t = \text{const}$ and $\vartheta = \pi/2$, the embedding function $\widetilde{z}(\xi)$, which characterizes the spatial profile of the wormhole, satisfies the differential equation \cite{Morris1988AJP}
\begin{equation}\label{embedding}
  \frac{d\widetilde{z}}{d\xi} = \pm \left( \frac{\xi}{b} - 1 \right)^{-1/2}, \quad \widetilde{z}(1) = 0.
\end{equation}
This surface, $\widetilde{z} = \widetilde{z}(\xi)$, is illustrated in Figure~\ref{profileb}, and \eqref{embedding} demonstrates how the shape function $b = b(\xi)$ fixes the spatial geometry of the wormhole. In Figure~\ref{profileb}, we also observe that the embedded surface corresponding to the non-commutative wormhole consistently lies within that of the Morris-Thorne wormhole. This reflects a less flared spatial geometry, i.e. a larger radius of curvature of the equatorial embedding surface near the throat, consistent with the geometry induced by the Gaussian‑smeared matter distribution.
\begin{table}[ht]
\caption{Representative examples of nontrivial redshift functions that are regular at the throat of the noncommutative geometry–inspired wormhole and vanish asymptotically at spatial infinity. The abbreviation `AP` denotes Adjustable Parameters, while $x_0$ marks the location of the throat. The parameters can be fine-tuned to ensure that $\Phi$ varies sufficiently slowly, thereby avoiding excessive tidal forces. The Type II redshift function exhibits a power-law decay of order $n$, offering a controllable asymptotic fall-off.}
\label{table:Phi}
\vspace*{1em}
\centering
\begin{tabular}{||c|l|c|c|c||}
\hline\hline
\text{Type} & \text{AP} & $\Phi(x)$ & $\Phi(x_0)$ & $\Phi_1=\Phi'(x_0)$ \\[0.5ex]
\hline\hline
I   & $\Phi_0$ 
    & $\Phi_0 e^{-(x-x_0)}$ 
    & $\Phi_0$ 
    & $-\Phi_0$ \\
II  & $\Phi_0,\ n\!\ge\!1,\ k\!>\!0$ 
    & $\frac{\Phi_0}{\bigl[1+(x-x_0)^n\bigr]^k}$ 
    & $\Phi_0$ 
    & $\begin{cases}
       -k\,\Phi_0, & \text{if } n=1,\\[2pt]
       0,          & \text{if } n>1,
       \end{cases}$ \\
III & $\Phi_0$ 
    & $\Phi_0\!\left[1-\tanh{(x-x_0)}\right]$ 
    & $\Phi_0$ 
    & $-\Phi_0$ \\
IV  & $\Phi_0$ 
    & $\Phi_0\,e^{-(x-x_0)^2}$ 
    & $\Phi_0$ 
    & $0$ \\[1ex]
\hline\hline
\end{tabular}
\end{table}
For an anisotropic fluid with energy density $\rho$, radial pressure $p_r$, and tangential pressure $p_t$, the NEC requires
\begin{equation}
T_{\mu\nu} k^\mu k^\nu \geq 0 
\end{equation}
for all null vectors $k^\mu$. In a local orthonormal frame adapted to the fluid \cite{Morris1988AJP}, this implies in particular
\begin{equation}
\rho + p_r \geq 0, \qquad \rho + p_t \geq 0 .
\end{equation}
Traversable wormhole geometries that satisfy the usual flare‑out condition at the throat necessarily violate at least one of these inequalities. Matter sources that produce $\rho + p_r < 0$ or $\rho + p_t < 0$ in some region are therefore referred to as exotic matter, in the sense that they violate the classical energy conditions used in standard singularity theorems. In our noncommutative wormhole model, we quantify the exoticity by the sign of $\rho + p_r$. Regions where $\rho + p_r < 0$ correspond to NEC‑violating exotic matter that supports the throat. In our present problem, we find
\begin{equation}\label{eq:NEC-general}
  \rho(x) + p_r(x) = \frac{1}{8\pi} \left(\frac{x b^{'} - b}{x^3} + \frac{2f}{x}\Phi^{'}\right),
\end{equation}
where a prime denotes differentiation with respect to the rescaled variable $x$. Evaluating this expression at the throat where $b(x_0)=x_0$ and $f$ vanishes,
\begin{equation}\label{NEC}
  \rho(x_0) + p_r(x_0) = \frac{b^{'}(x_0) - 1}{8\pi x_0^2} < 0,
\end{equation}
as it can be evinced from Table~\ref{table:throat}. The negativity of $\rho + p_r$ at the throat reflects a violation of the NEC and therefore signals the presence of exotic matter required to sustain the flare–out geometry. This is consistent with the general behaviour of traversable wormholes, where at least one of the null energy condition inequalities must fail in a neighbourhood of the throat. In addition to the NEC, it is useful to recall the standard pointwise energy conditions in an orthonormal frame adapted to the fluid. For the anisotropic stress–energy tensor \eqref{emt}, the weak energy condition (WEC) requires $\rho \geq 0$, $\rho + p_r \geq 0$, $\rho + p_t \geq 0$, the dominant energy condition (DEC) further demands $\rho \geq |p_r|$, $\rho \geq |p_t|$, and the strong energy condition (SEC) is usually stated as $\rho + p_r + 2 p_t \geq 0$ together with the NEC. In our noncommutative wormhole model, the Gaussian-smeared shape function implies a strictly positive energy density, $\rho>0$, everywhere (see \eqref{density}). Consequently, violations of the NEC near the throat automatically entail violations of the WEC and the SEC in the same region. For the representative redshift families analysed in Section II (see Figures 3-6), one can obtain configurations where $\rho+p_r$ becomes nonnegative over an interval away from the throat. In the EOS-based constructions of Section III, the matter sector is engineered to approach the de Sitter-like regime $p_r\simeq-\rho$ at large radii, so that $\rho+p_r\to 0$ asymptotically. Last but not least, notice that the NEC saturates for $x=x_e$. Thus, if one looks only at the throat value, the impact of the redshift might appear limited. However, away from the throat, it controls the stress–energy gradients, tidal forces, time dilation, and, crucially for this work, the localisation and magnitude of NEC violation. Imposing $\rho+p_r\geq 0$ for $x>x_0$ requires 
\begin{equation}\label{eq:Phi-ineq}
\Phi^{'}\geq\frac{b-xb^{'}}{2x(x-b)}.    
\end{equation}
Near the throat, the above inequality becomes
\begin{equation}
\Phi^{'}\gtrsim\frac{1}{2(x-x_0)}+\mathcal{O}(1),   
\end{equation}
and its right-hand side diverges as $x\to x_0^+$. So a regular redshift, i.e. $\Phi^{'}$ finite, cannot make the NEC nonnegative arbitrarily close to $x_0$. A NEC‑violating layer is unavoidable, though its thickness can be minimised. To quantify how the redshift function affects the near‑throat behaviour, let $b^{'}(x_0)=b_1$, $b^{''}(x_0)=b_2$, and $\Phi^{'}(x_0)=\Phi_1$. From Table~\ref{table:throat}, we have $0<b_1<1$ and $-1<b_2<0$. A straightforward differentiation of \eqref{eq:NEC-general} gives the radial slope at the throat
\begin{equation}\label{eq:NEC-slope}
  8\pi\left.\frac{d}{dx}(\rho+p_r)\right|_{x=r_0}=\frac{b_2}{x_0^2}+\frac{3}{x_0^3}(1-b_1)+\frac{2}{x_0^2}(1-b_1)\Phi_1.   
\end{equation}
Since the factor $1-b_1$ is always positive (see Table~\ref{table:throat}), increasing $\Phi_1$ raises $8\pi(\rho+p_r)$ just outside the throat, in other words, the NEC violation is reduced. When a near‑throat zero $x_c>x_0$ of $8\pi(\rho+p_r)$ exists, a first‑order Taylor estimate for the width of the violating layer is
\begin{equation}\label{eq:rc-deriv}
  x_c-x_0\approx\frac{1-b_1}{b_2+(1-b_1)\left(2\Phi_1+\frac{3}{x_0}\right)}. 
\end{equation}
Whenever the denominator is positive, this width decreases monotonically with $\Phi_1$
\begin{equation}
  \frac{\partial}{\partial\Phi_1}(x_c-x_0)\approx-\frac{2(1-b_1)^2}{\left[b_2+(1-b_1)\left(2\Phi_1+\frac{3}{x_0}\right)\right]^2}<0.
\end{equation}
Hence, increasing $\Phi_1$ shrinks the violating shell, and this behaviour is independent of the sign of $b_2$. Furthermore, the condition for $x_c>x_0$ in the linear approximation is
\begin{equation}\label{eq:rc-positivity}
  \Phi_1\gtrsim-\frac{1}{2}\left(\frac{3}{x_0}+\frac{b_2}{1-b_1}\right).    
\end{equation}
We underline the fact that equations \eqref{eq:Phi-ineq}–\eqref{eq:rc-positivity} are model–independent. They show that the redshift potential does not affect the sign of the NEC exactly at the throat, but it governs how quickly the NEC recovers, and how thin the violating layer can be. To turn these inequalities into constructive wormhole models, one still needs explicit, horizon–avoiding redshift profiles that
\begin{enumerate}[label=(\roman*), leftmargin=*]
\item are regular at the throat ($|\Phi|,|\Phi^{'}|<\infty$ and $e^{2\Phi(x_0)}\neq 0$),
\item ensure asymptotic flatness ($\Phi\to 0$ as $x\to+\infty$),
\item satisfies the design target of getting the NEC restored beyond a thin shell,
\item and sample distinct near‑throat gradients and asymptotics so that the trends implied by \eqref{eq:NEC-slope}–\eqref{eq:rc-deriv} can be seen in concrete spacetimes.
\end{enumerate}
For these reasons, we complement the general bounds with a small set of representative redshift profiles (Types I–VI in Table~\ref{table:Phi}). They are not arbitrary handpicked functions because together they span the key features that matter here, namely the sign/magnitude of $\Phi_1$, the presence or absence of a local bump which confines the gradient, and the asymptotic falloff, i.e. power‑law vs.\ exponential. To make the impact of different redshift potentials on the energy conditions more transparent, we now discuss the four representative redshift families of Table~\ref{table:Phi} as
separate models, focusing in each case on the behaviour of $p_r$, $p_t$, and the NEC.

\begin{figure}[!ht]
\centering
    \includegraphics[width=0.3\textwidth]{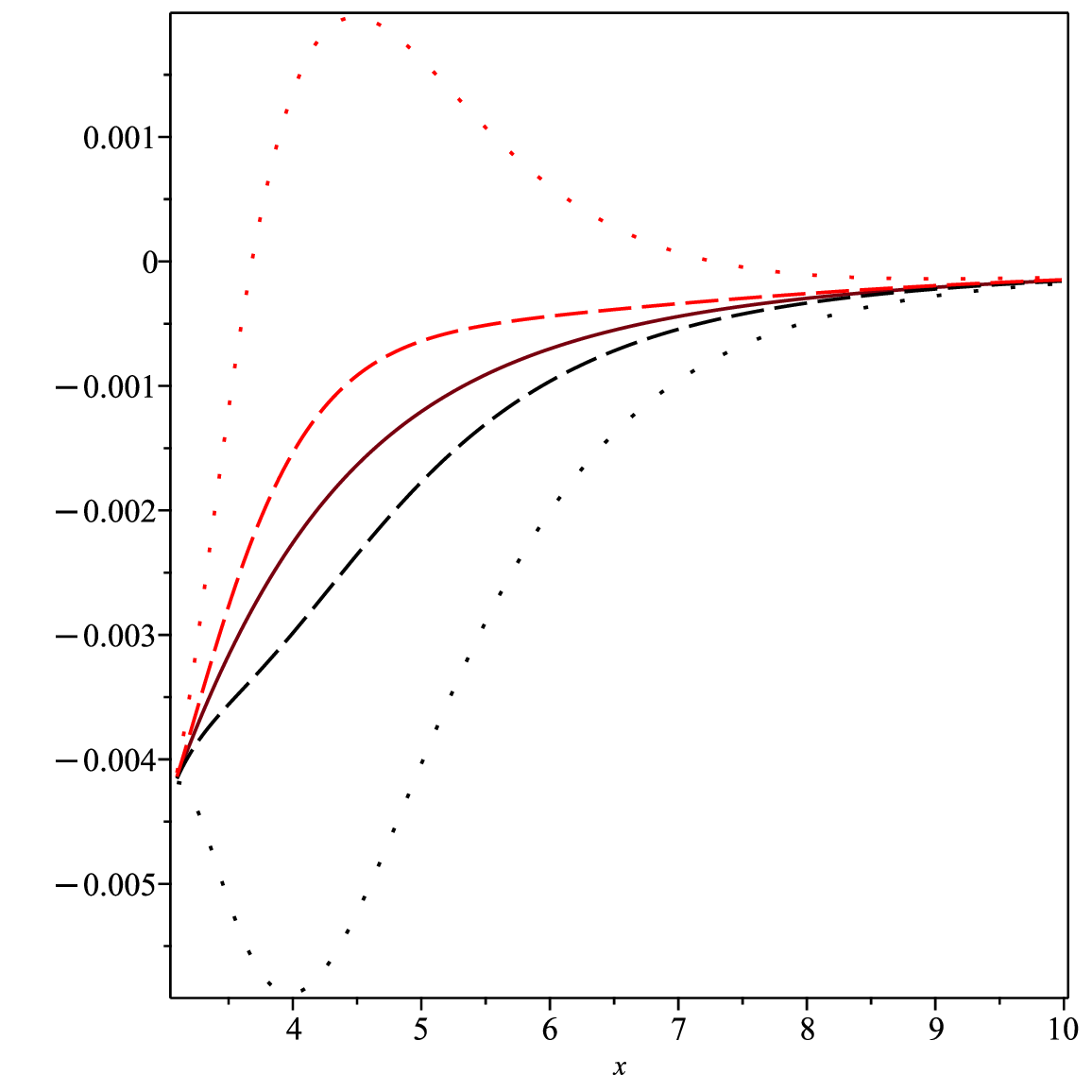}
    \includegraphics[width=0.3\textwidth]{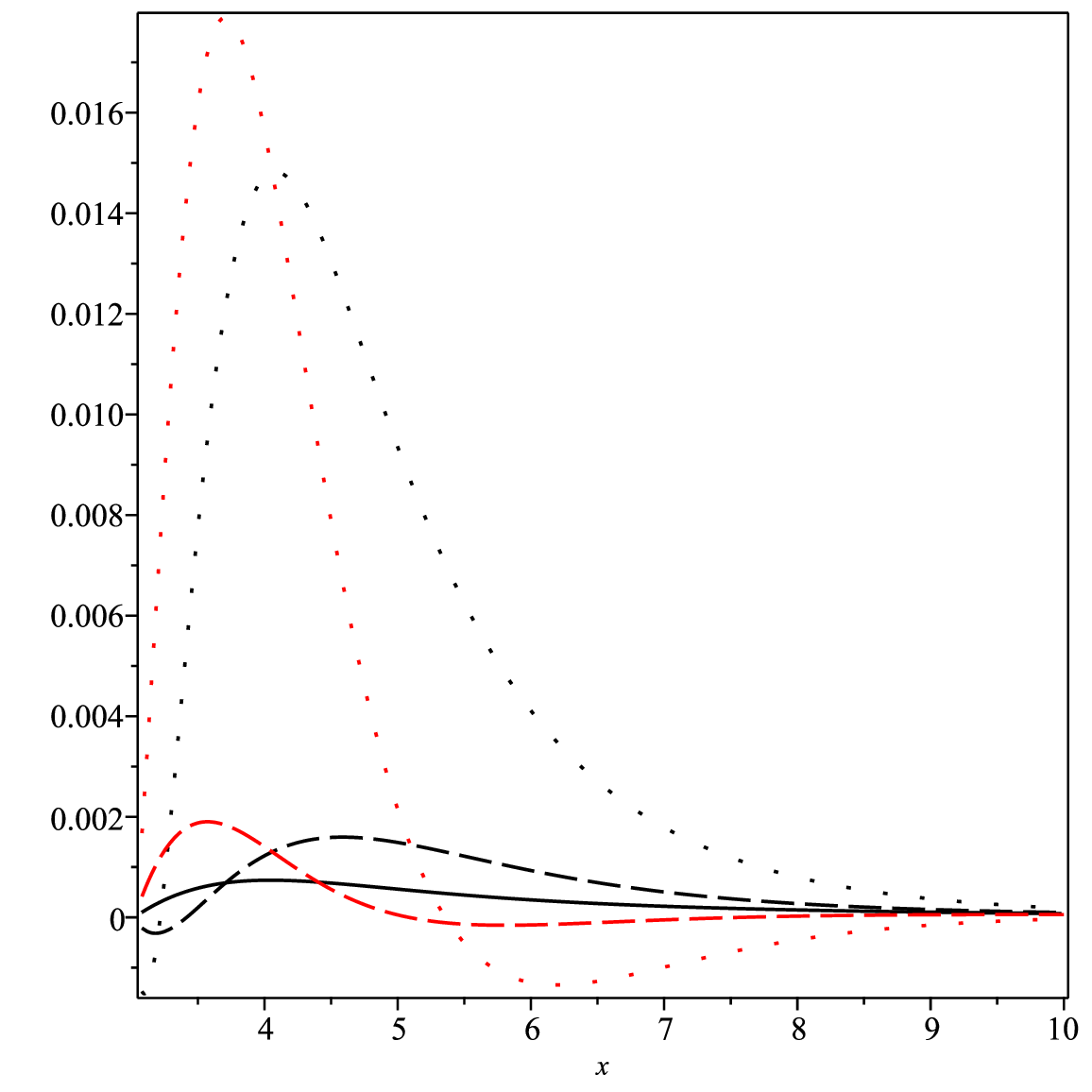}
    \includegraphics[width=0.3\textwidth]{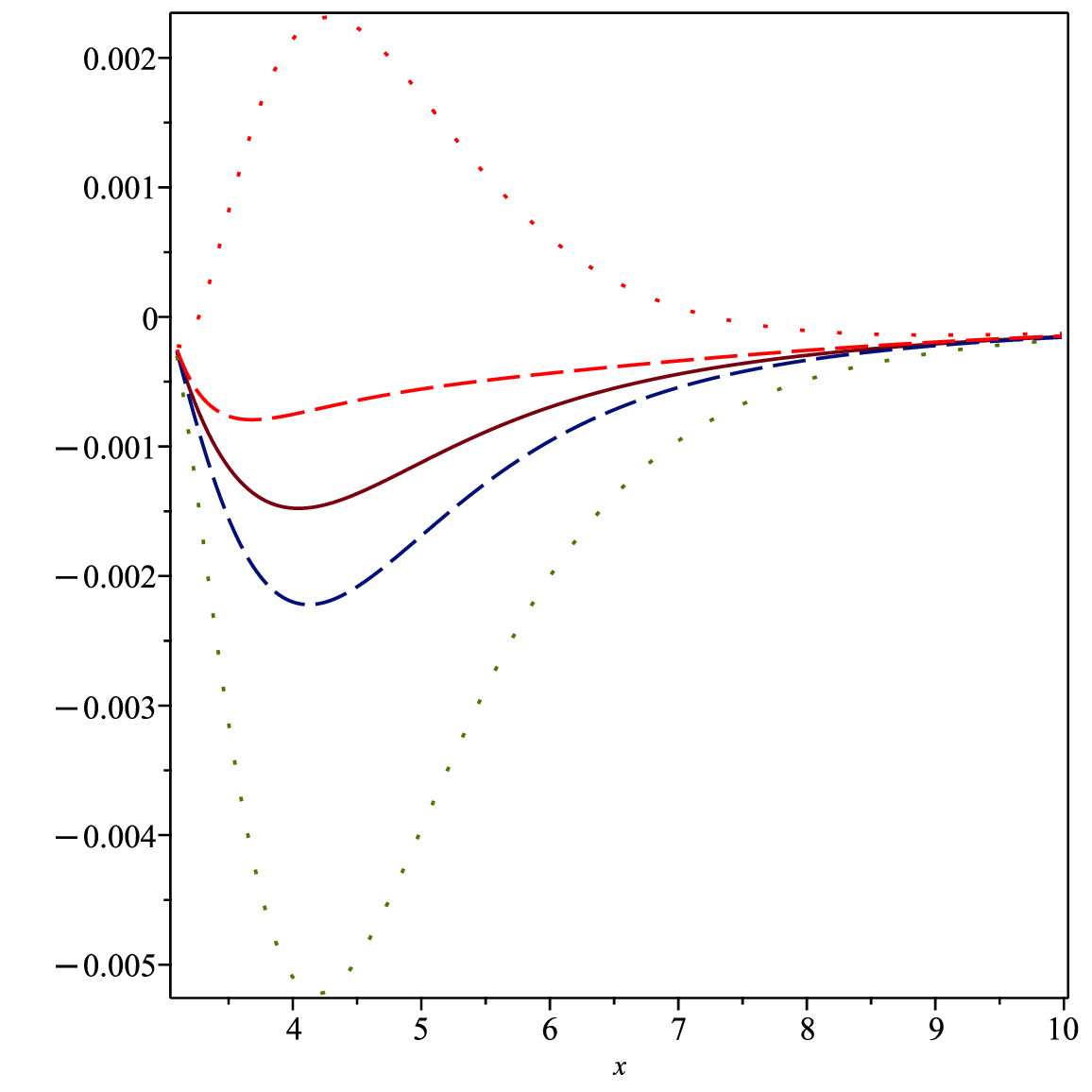}
\caption{\label{PhiI} {\bf{Left panel}}: Radial pressure $p_r$ plotted as a function of the rescaled radial variable $x$ for $\mu = 1.905$, and various values of $\Phi_0$ in the type I redshift function $\Phi$ (see Table~\ref{table:Phi}). The case $\Phi_0 = 0$ is shown as a solid line; $\Phi_0 = 1$ as a black dashed line; $\Phi_0 = 5$ as a black spacedotted line; $\Phi_0 = -1$ as a red dashed line; and $\Phi_0 = -5$ as a red spacedotted line. {\bf{Central panel}}: Tangential pressure $p_t$ plotted against $x$ for the same set of parameters and line styles as in the left panel. {\bf{Right panel}}: NEC represented against $x$ for the same set of parameters and line styles as in the left panel.}
\end{figure}

\begin{figure}[!ht]
\centering
    \includegraphics[width=0.3\textwidth]{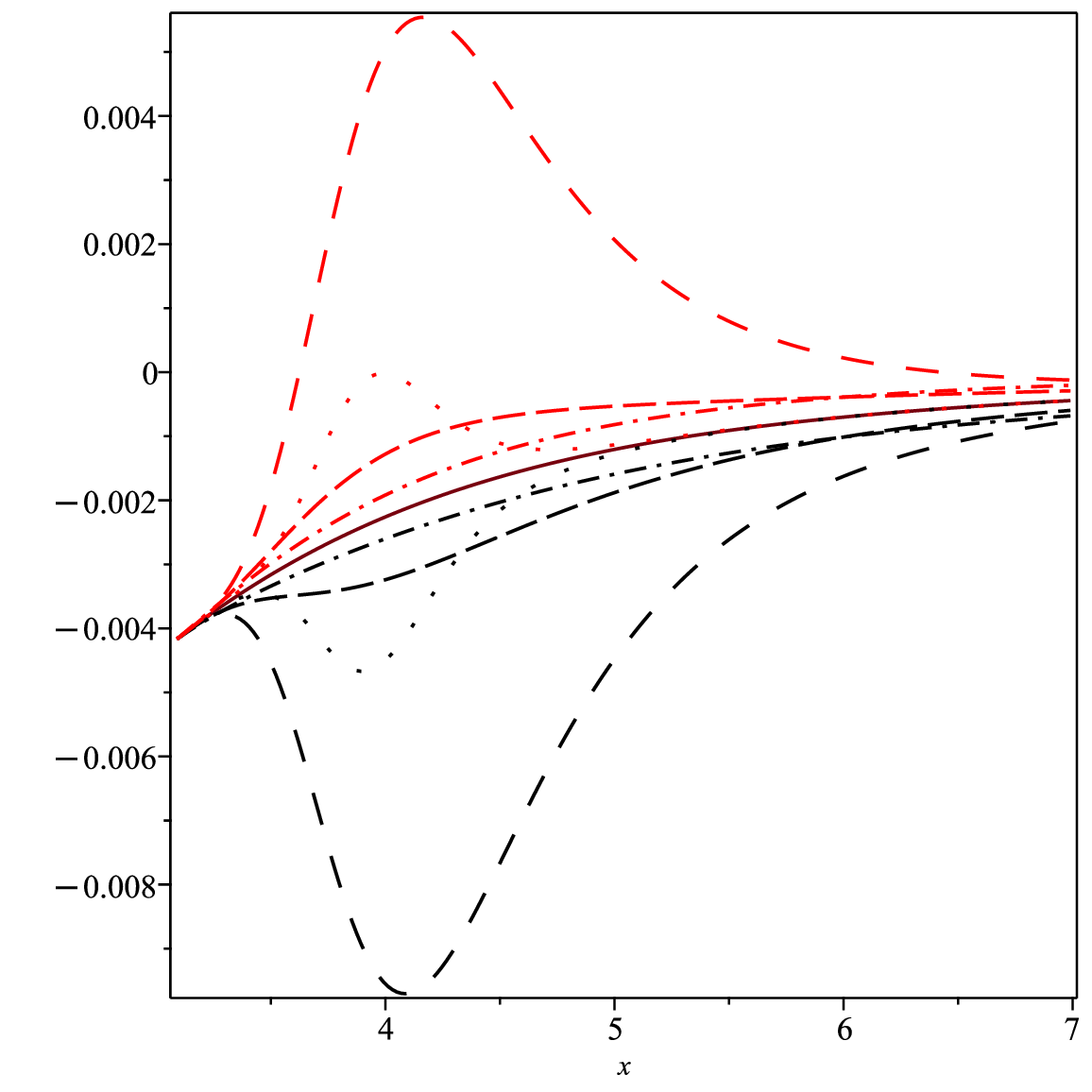}
    \includegraphics[width=0.3\textwidth]{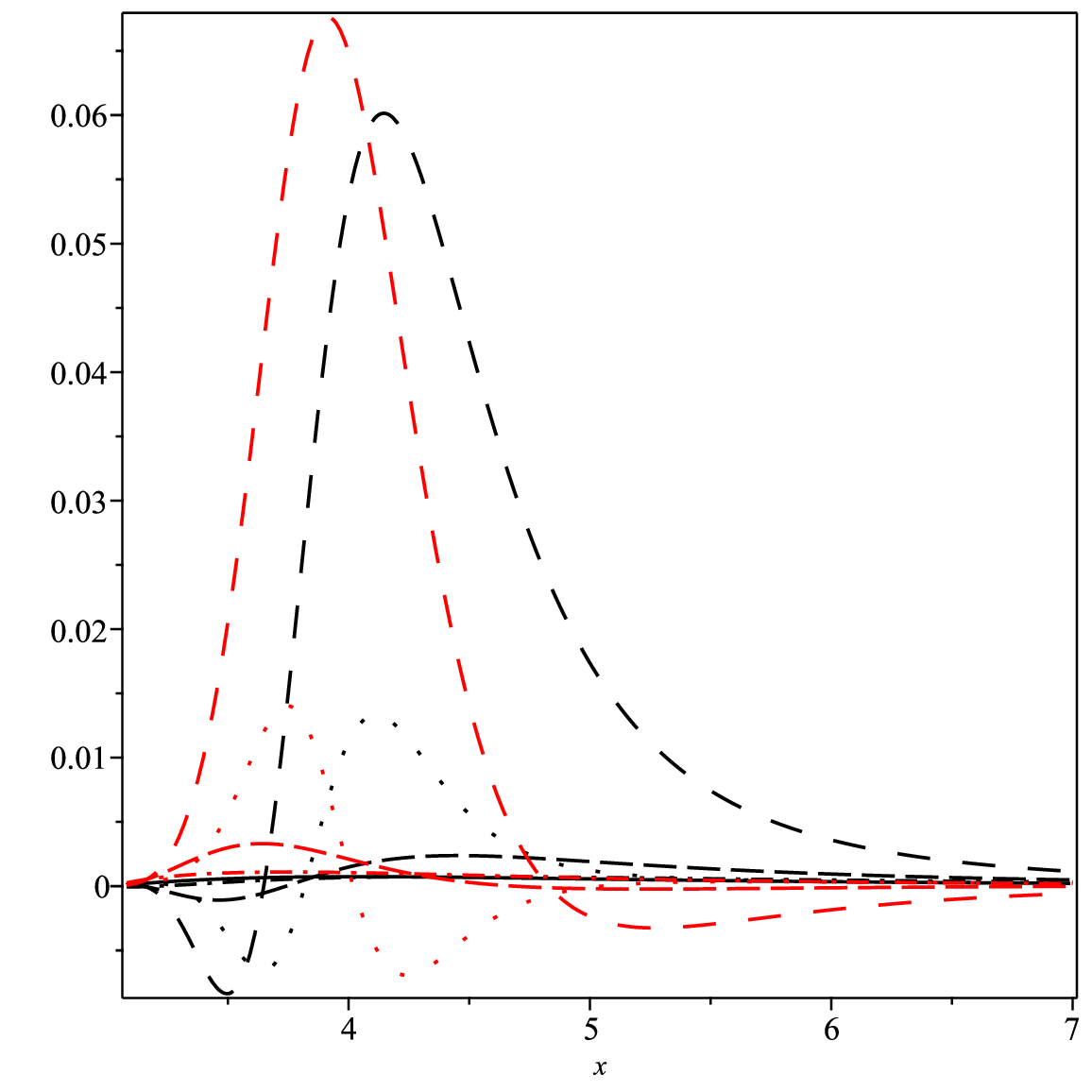}
    \includegraphics[width=0.3\textwidth]{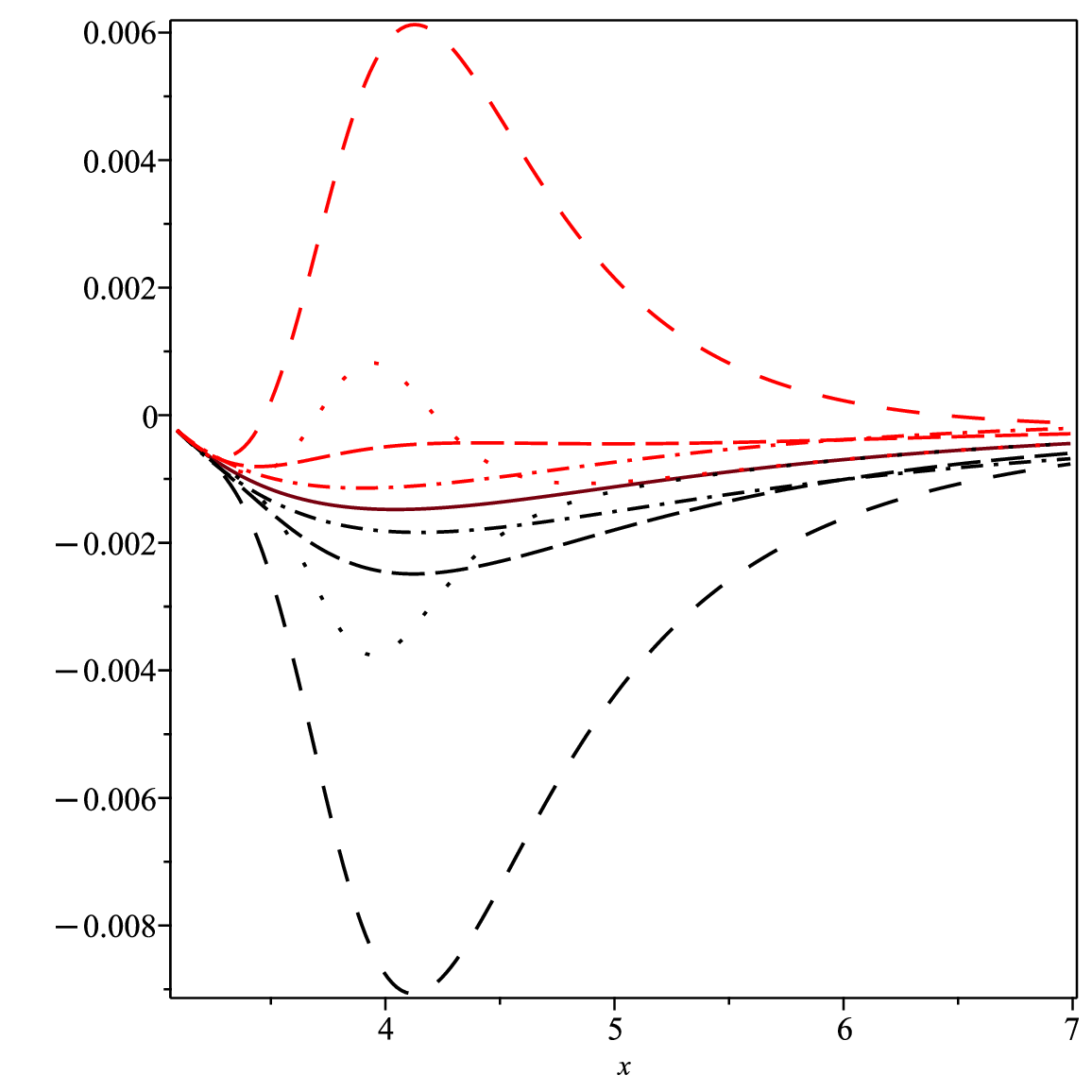}
\caption{\label{PhiII} {\bf{Left panel}}: Radial pressure $p_r$ plotted as a function of the rescaled radial variable $x$ for $\mu = 1.905$, and various parameter triples $(\Phi_0, n, k)$ used in the type II redshift function $\Phi$ (see Table~\ref{table:Phi}). The reference case $\Phi_0 = 0$ is shown as a solid line. The triple $(\pm 1, 2, 1)$ is represented by dashed lines (positive $\Phi_0$ in red), $(\pm 1, 4, 2)$ by spacedotted lines (positive in red), $(\pm 1, 1, 1/2)$ by dash-dotted lines (positive in red), and $(\pm 5, 3, 1)$ by spacedashed lines (positive in red). {\bf{Central panel}}: Tangential pressure $p_t$ plotted against $x$ for the same parameters, triples, and line styles as in the left panel. {\bf{Right panel}}: NEC represented against $x$ for the same set of triples and line styles as in the left panel.}
\end{figure}

\begin{figure}[!ht]
\centering
    \includegraphics[width=0.3\textwidth]{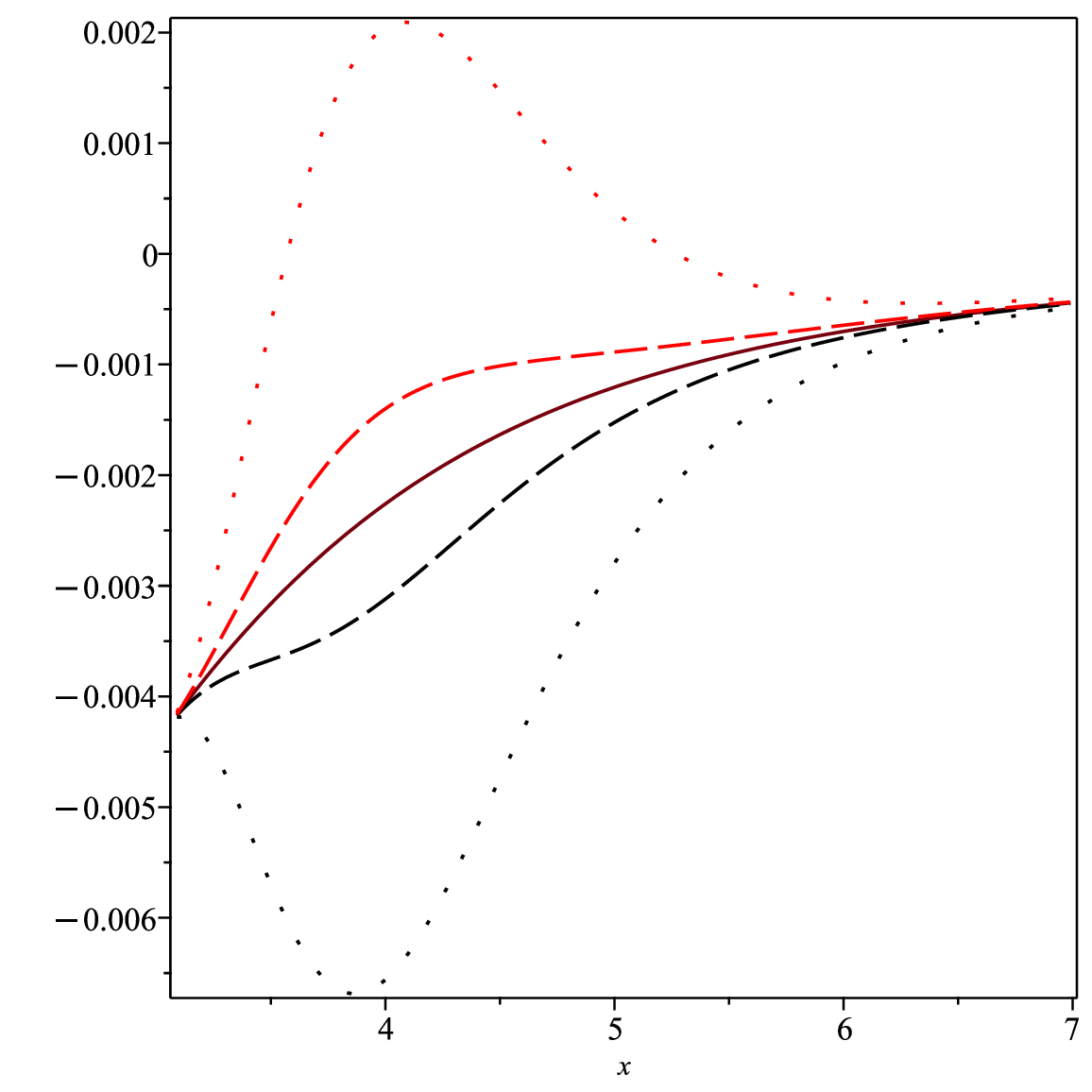}
    \includegraphics[width=0.3\textwidth]{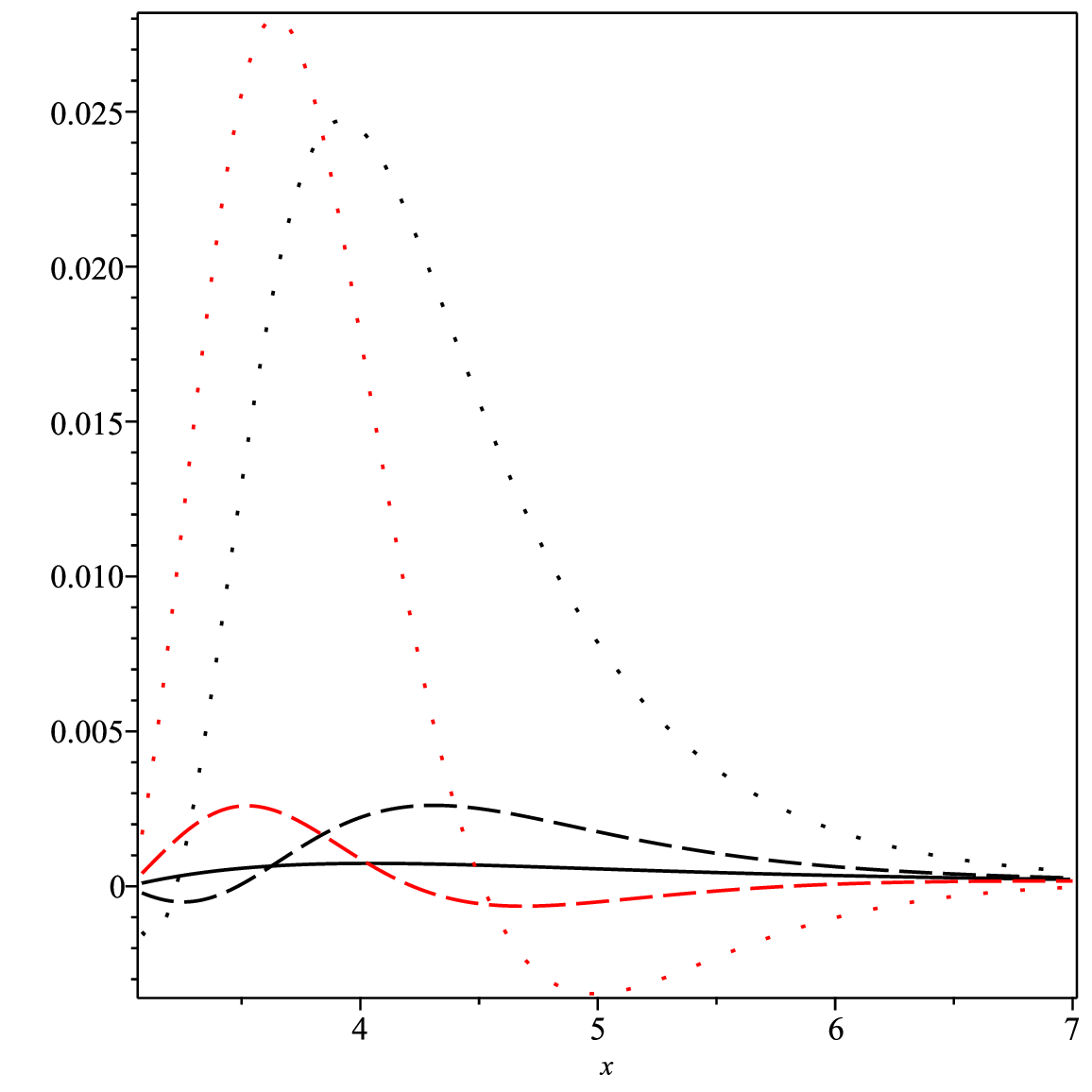}
    \includegraphics[width=0.3\textwidth]{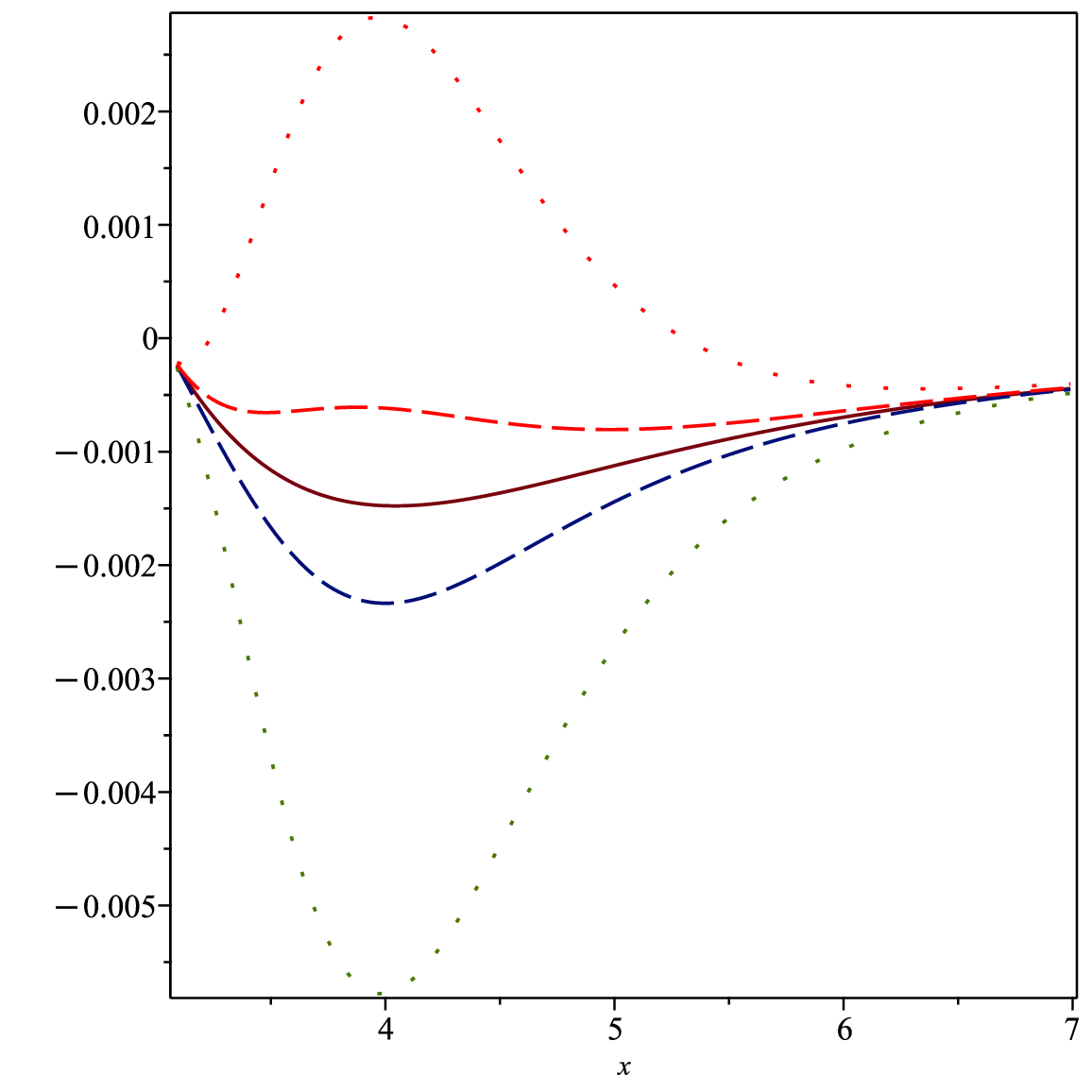}
\caption{\label{PhiIII} {\bf{Left panel}}: Radial pressure $p_r$ plotted as a function of the rescaled radial variable $x$ for $\mu = 1.905$, and various values of $\Phi_0$ used in the type III redshift function $\Phi$ (see Table~\ref{table:Phi}). The case $\Phi_0 = 0$ is shown as a solid line, $\Phi = 1$ as a dashed line, $\Phi_0 = 5$ as a spacedotted line, $\Phi_0 = -1$ as a red dashed line, and $\Phi_0 = -5$ as a red spacedotted line. {\bf{Central panel}}: Tangential pressure $p_t$ plotted against $x$ for the same choices of $\Phi_0$ and line styles as in the left panel. {\bf{Right panel}}: NEC represented against $x$ for the same set of values of $\Phi_0$ and line styles as in the left panel.}
\end{figure}

\begin{figure}[!ht]
\centering
    \includegraphics[width=0.3\textwidth]{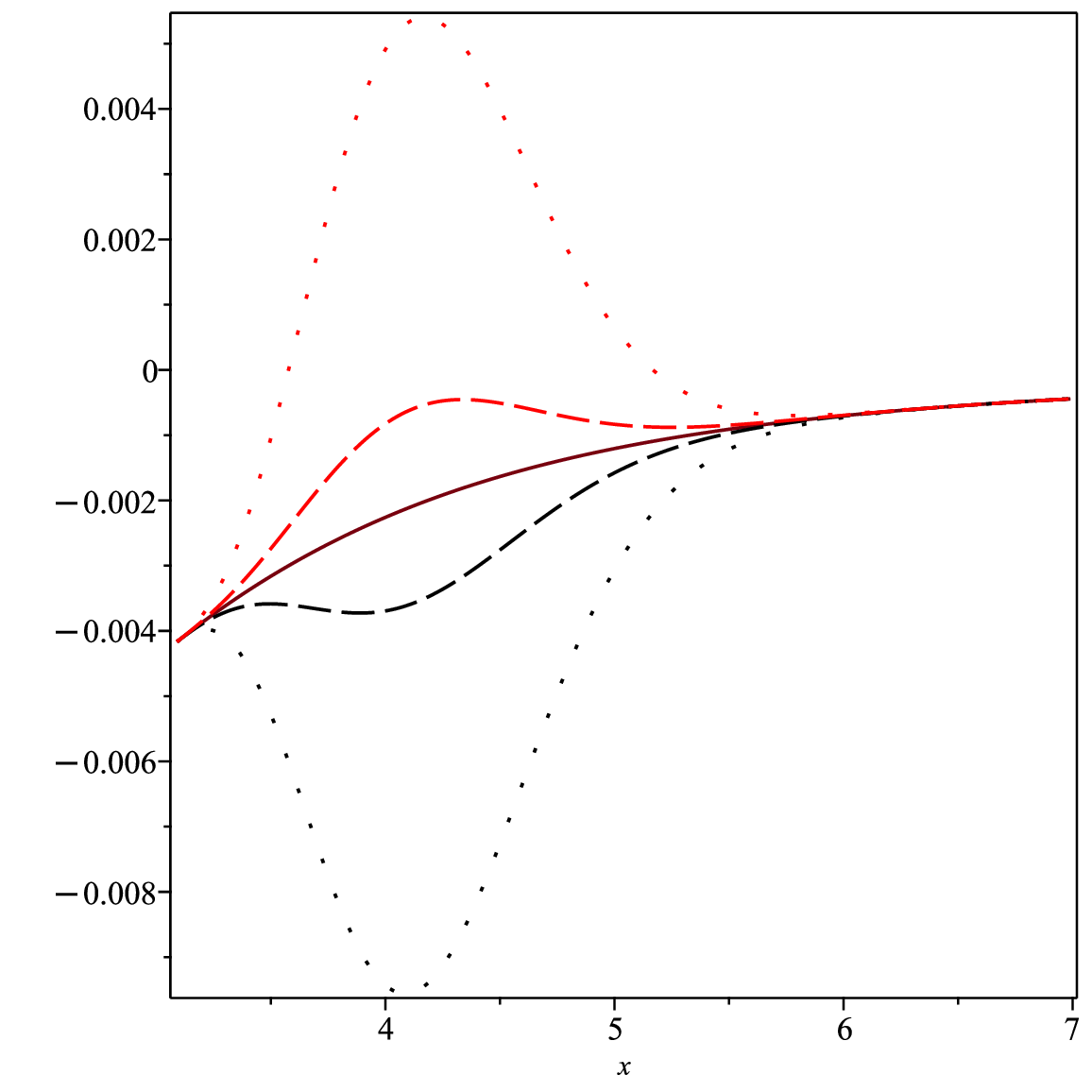}
    \includegraphics[width=0.3\textwidth]{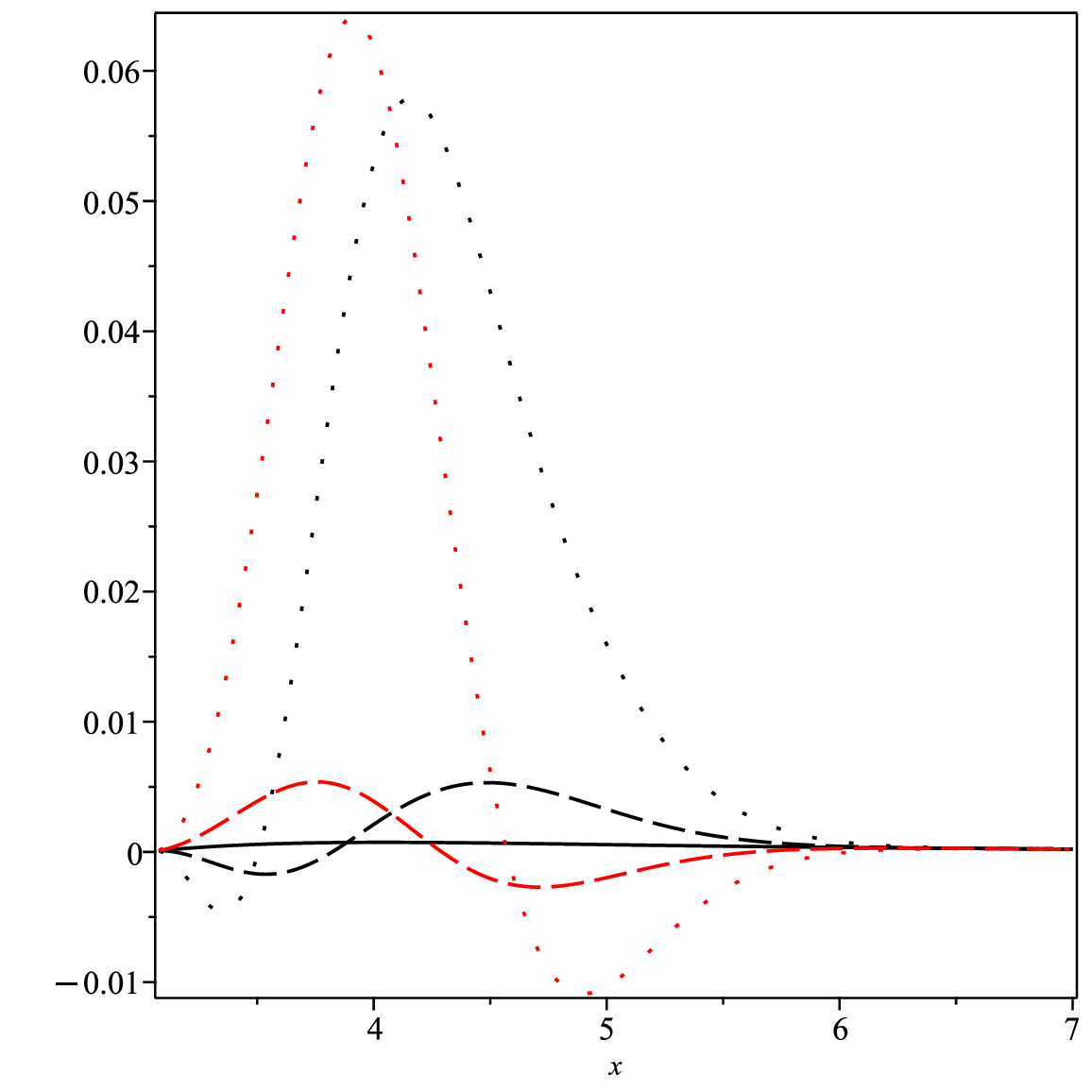}
    \includegraphics[width=0.3\textwidth]{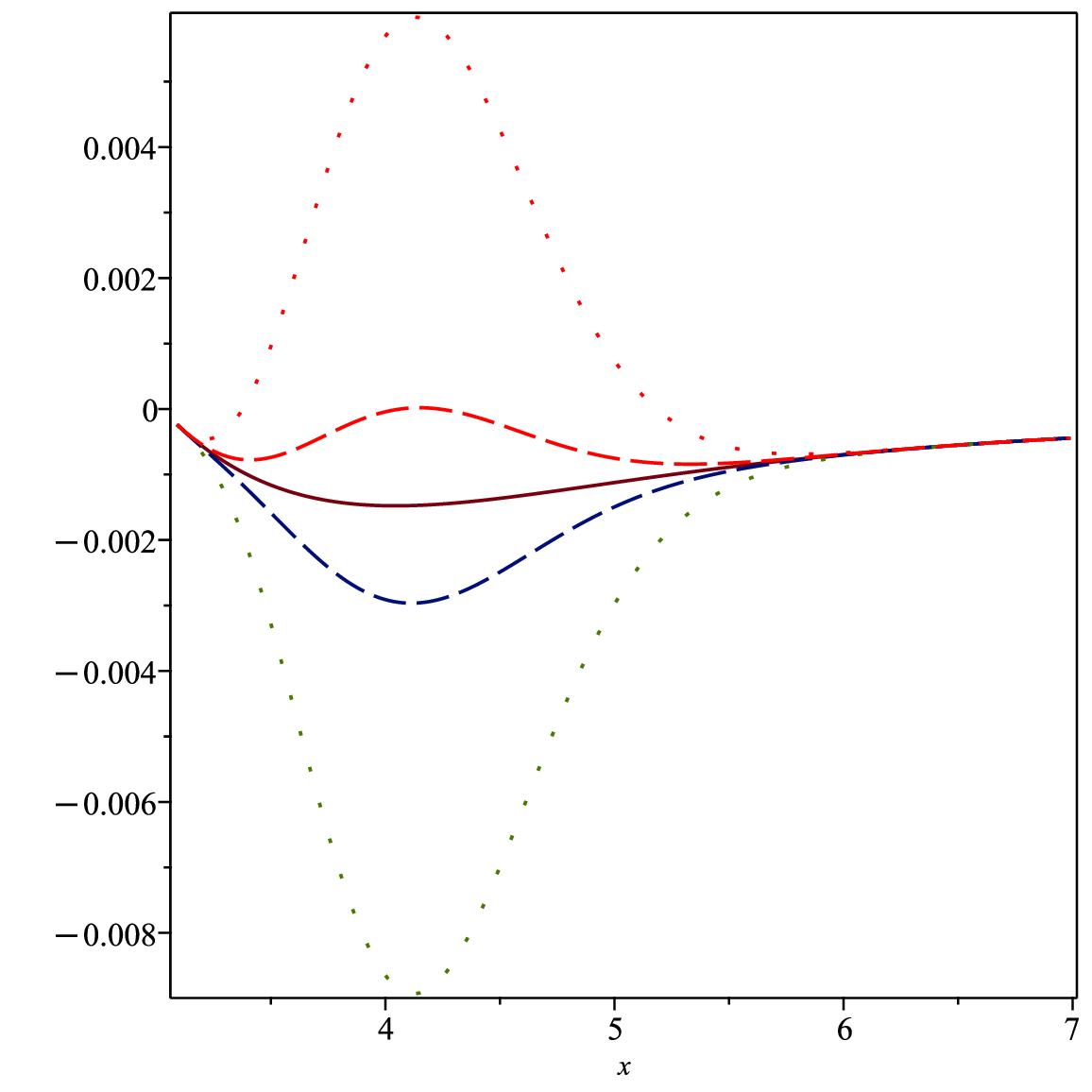}
\caption{\label{PhiIV} {\bf{Left panel}}: Radial pressure $p_r$ plotted as a function of the rescaled radial variable $x$ for $\mu = 1.905$, and various values of $\Phi_0$ used in the type IV redshift function $\Phi$ (see Table~\ref{table:Phi}). The case $\Phi_0 = 0$ is shown as a solid line, $\Phi_0 = 1$ as a dashed black line, $\Phi_0 = 5$ as a black spacedotted line, $\Phi_0 = -1$ as a red dashed line, and $\Phi_0 = -5$ as a red spacedotted line. {\bf{Central panel}}: Tangential pressure $p_t$ plotted against $x$ for the same choices of $\Phi_0$ and line styles as in the left panel. {\bf{Right panel}}: NEC represented against $x$ for the same set of values of $\Phi_0$ and line styles as in the left panel.}
\end{figure}

\subsection{Model I: exponential redshift}

Let us first consider the redshift $\Phi(x)=\Phi_0 e^{-(x-x_0)}$ (see Table~\ref{table:Phi}), so the near‑throat gradient is $\Phi_1=-\Phi_0$. Thus, a negative $\Phi_0$ corresponds to a positive $\Phi_1$, which tends to lift the NEC just outside the throat, whereas a positive $\Phi_0$, i.e. $\Phi_1<0$, pushes the NEC downward and deepens the violation. This trend is clearly visible in Fig.~\ref{PhiI}. For $\Phi_0>0$, the radial pressure $p_r$ becomes increasingly negative, indicating stronger radial tension. For $\Phi_0<0$, the tension relaxes and $p_r$ can turn positive over a finite interval (e.g., for $\Phi_0 = -5$, a positive peak appears at $x_p\simeq 4.4815$). The tangential pressure $p_t$ shows a complementary behaviour. When $\Phi_0>0$, it dips below zero near the throat, while for $\Phi_0<0$ it remains positive close to $x_0$ and only later turns negative. In all cases, $p_t$ exhibits a local maximum that moves inward as $|\Phi_0|$ increases, indicating enhanced anisotropic support near the throat. For instance, the maxima lie at $x_m\simeq 3.182$ and $3.098$ for $\Phi_0\in\{1,5\}$, and at $x_m\simeq 3.572$ and $3.708$ for $\Phi_0\in\{-1,-5\}$, respectively. The NEC plots mirror this pattern. Positive $\Phi_0$ enlarges the region where the NEC is violated, while negative $\Phi_0$ both weakens and localises the violation. For $\Phi_0=-5$, the first zero occurs at $x_c\simeq 3.2674$, the NEC then stays positive up to $x\simeq 7.2710$ before turning negative again only in the far tail. This is precisely the localisation effect anticipated by the near‑throat control parameter $\Phi_1$. A brief comment on thresholds helps interpret these curves. For the case $\mu=1.905$, our near‑throat condition for immediate NEC recovery gives $\Phi_1>8.21$, which translates to $\Phi_0\lesssim-8.21$. Yet the first zero of the NEC already appears numerically for $\Phi_0=-5$. There is no contradiction. The inequality \eqref{eq:rc-positivity} is a sufficient condition that guarantees the NEC starts increasing right at the throat. It is not necessary for a crossing to occur at a finite radius. In terms of the Taylor expansion of the NEC about the throat, the linear term sets that sufficient bound, while the quadratic correction can still drive the NEC to zero slightly away from the throat even when the linear criterion is not met. Including the quadratic approximation sharpens the estimate and is consistent with the crossing observed at $\Phi_0 = -5$.

\subsection{Model II: rational power-law decay}

Figure~\ref{PhiII} shows that the overall patterns of $p_r$, $p_t$, and the NEC in the case of redshifts of type II closely mirror those found for Type I. The reference case ($\Phi_0=0$, solid black curve) reproduces the classical noncommutative wormhole studied by \cite{Garattini2009PLB} where the radial pressure remains negative, the tangential pressure is always positive outside the throat, and the NEC is violated everywhere. Once $\Phi_0$ is switched on, the behaviour changes qualitatively. For example, with the triple $(\Phi_0,n,k)=(-5,3,1)$ the radial pressure crosses zero at $x\simeq 3.6241$, reaches a positive peak at $x_p\simeq 4.1647$, turns negative again for $x\gtrsim 6.4301$, attains a negative minimum at $x_m\simeq 7.9194$, and then decays toward zero. The tangential pressure displays the same near‑throat trends as in Type I. For the triple $(5,3,1)$, it is positive almost everywhere but dips below zero close to the throat ($x_0<x<3.1535$) and develops a peak at $x_p\simeq 4.1462$. For $(-5,3,1)$, it stays positive from the throat out to $x\simeq 4.8275$, exhibits a higher maximum at $x_p\simeq 3.9143$ than that detected for the triple $(5,3,1)$, and then shows a negative minimum at $x_m\simeq 5.2686$. The NEC follows the same logic. Positive $\Phi_0$ amplifies and radially extends the violation, while negative $\Phi_0$ mitigates and confines it. For $(-5,3,1)$, the first zero occurs at $x_c\simeq 3.4735$. The NEC then reaches a positive maximum at $x_p\simeq 4.1280$, becomes negative again for $x\gtrsim 6.4341$, attains a negative minimum at $x_m\simeq 7.9199$, and tends to zero at large $x$. At this point, a comment is in order. Notice that for redshifts of type II (see Table~\ref{table:Phi}), we indeed have $\Phi_1=0$ for all $n>1$. That means the redshift is flat at the throat, and its leading near‑throat influence enters through $\Phi_2=\Phi^{''}(x_0)$. The sign of $\Phi_0$ then fixes the sign of $\Phi_2$. For $n>1$, $x_0$ is a local maximum when $\Phi_0>0$ so $\Phi_2<0$, and a local minimum when $\Phi_0<0$ so $\Phi_2>0$. Because the near‑throat derivative of the NEC  depends on $\Phi_1$ but not on $\Phi_2$ because in the expression for the first derivative of $p_r+\rho$, the function $\Phi^{''}$ is multiplied by $f(x)$ which vanishes at $x=x_0$, \eqref{eq:NEC-slope} is unchanged by turning on $\Phi_0$ in Type II redshifts. $\Phi_2$ enters at the quadratic order and controls how quickly the NEC bends upward or downward just outside the throat. This is why Type II’s qualitative trends resemble Type I’s, but the onset of features, such as, for instance, the first zero of the NEC, occurs slightly farther than in the Type I case.

\subsection{Model III: sigmoidal/tanh profile}

In the case of redshifts of type III (see Figure~\ref{PhiIII}), for $\Phi_0=-5$, the radial pressure is negative from the throat up to $x\simeq 3.5711$, reaches a positive maximum at $x_p\simeq 4.0914$, crosses zero again at $x\simeq 5.2687$, develops a negative local minimum at $x_m\simeq 6.3430$, and then decays toward zero. The tangential pressure shows the expected near‑throat behaviour seen in earlier families, but with a sigmoidal imprint. For $\Phi_0=5$, it dips below zero only close to the throat ($x_0<x<3.2250$), peaks at $x_p\simeq 3.9231$, and then remains positive as it relaxes to zero. For $\Phi_0=-5$, it is non‑negative up to $x\simeq 4.5160$, exhibits a higher positive maximum at $x_p\simeq 3.6439$, then a negative minimum at $x_m\simeq 4.9880$, before tending to zero at large $x$. The NEC follows the same logic as in Type I. Positive $\Phi_0$ amplifies and radially extends the region where the NEC is violated, while negative $\Phi_0$ mitigates and localises it. For $\Phi_0=-5$, the violation is confined to a near‑throat interval $x_0<x<3.2360$. The NEC then climbs to a positive maximum at $x_p\simeq 3.9666$, crosses zero again at $x\simeq 5.3028$, dips to a small negative minimum at $x_m\simeq 6.3593$, and finally approaches zero from below. Overall, the sigmoidal redshift function preserves the qualitative trends dictated by the near‑throat control parameter $\Phi_1$, while shifting the locations of the features in a way consistent with its smooth, step‑like decay.

\subsection{Model IV: Gaussian redshift}

For Gaussian redshifts (see type IV in Table~\ref{table:Phi}), the leading near‑throat influence enters at quadratic order through $\Phi_2$. As a result, the immediate throat value of $8\pi(\rho+p_r)$ is unchanged, but negative $\Phi_0$ bends the NEC upward just outside the throat, thus mitigating and localising the violation, whereas positive $\Phi_0$ bends it downward, thus amplifying and extending it (see Figure~\ref{PhiIV}). In line with this, the radial and tangential pressures exhibit the same qualitative trends seen in Type I, with feature locations shifted slightly away from the throat due to the vanishing of $\Phi_1$. For illustration, with $\Phi_0=-5$, the NEC is violated only in a near‑throat interval $x_0<x<3.3766$. Then it rises to a positive maximum at $x\simeq4.1312$, crosses zero again at $x\simeq5.1850$, dips to a negative minimum at $x\simeq5.8097$, and finally relaxes towards zero at larger $x$.

Across the four families of redshift functions considered in the present work, the sign of the parameter $\Phi_0$ emerges as the primary control parameter.  Positive values of $\Phi_0$ deepen the gravitational well and enhance the need for exotic matter. In contrast, negative values tend to relax the energy conditions and can even yield regions in which the NEC is satisfied. The most economical configurations, which require the least violation of the NEC, arise when negative values of $\Phi_0$ are combined with redshift profiles that flatten asymptotically and descend rapidly (as in type III). Importantly, these optimisations are achieved without the formation of horizons while maintaining asymptotic flatness, thereby preserving the physical viability of the wormhole solutions.

\section{Wormholes Supported by Quasi-de Sitter and Chaplygin-Like EOS}\label{EOSmaster}

In the previous section, we left the EOS unspecified. Accordingly, the $tt$–field equation fixed the shape function $b$, while the redshift $\Phi$ had to be prescribed as an external profile. Once an EOS is chosen, however, the remaining Einstein equations close the system, and $\Phi$ is uniquely determined by $b$ and the chosen matter law. A natural first test is the strict de Sitter–like EOS $p_r=-\rho$, as in noncommutative geometry–inspired black holes \cite{Nicolini2006PLB}. Imposing $p_r=-\rho$ all the way to the throat and requiring a regular redshift function forces $b'(x_0)=1$, which together with the throat identity $8\pi(\rho+p_r)|_{x_0}=[b'(x_0)-1]/x_0^2$ saturates the flare‑out condition and yields a marginal (extremal), non‑traversable configuration. Thus, a traversable wormhole necessarily requires a localised departure from $p_r=-\rho$. This point is already visible in the simplest constant redshift case $\Phi=0$, which fixes the anisotropy through the EOS \eqref{pteos}. In other words, choosing $\Phi$ (e.g. $\Phi=0$) implicitly selects an EOS. To obtain a horizon‑free wormhole, one must allow a controlled, localised deviation from the pure de Sitter relation near the throat. To model such a scenario, we allow for a positive radial deviation
\begin{equation}\label{perturbation}
  p_r(x)=-\rho(x)\left[1+\delta(x)\right], \qquad \delta(x)>0~\forall x\geq x_0\,
\end{equation}
taking $\delta$ to be a localised Gaussian or Lorentzian bump that vanishes for $x\gg 1$. As we will see, $\delta$ can also be chosen to construct a Chaplygin-inspired EOS.  The function $\delta(x)$ encodes a localised departure from a de Sitter EOS. Since $\delta(x)>0$, one has $\rho+p_r=-\rho\delta(x)<0$ for all $x\ge x_0$, so in this quasi-de Sitter sector the NEC is not restored pointwise at a finite radius; rather, the NEC is approached asymptotically, $\rho+p_r\to 0^{-}$ as $x\to\infty$. In what follows, localisation refers to the fact that $\delta(x)$, and thus $|\rho+p_r|$, decays rapidly away from the throat, making the NEC violation negligible outside a narrow neighbourhood of $x_0$. In particular, we look for simple, positive, even profiles of the rescaled variable $x-x_0$ that (i) peak at the throat, (ii) vanish at infinity, and (iii) keep $\Phi$ regular via the condition \eqref{flare}. The two canonical choices we adopt
\begin{equation}\label{deltaGL}
\delta_{G}(x)=\delta(x_0)\,e^{-(x-x_0)^2}
\quad\text{and}\quad
\delta_{L}(x)=\frac{\delta(x_0)}{1+(x-x_0)^2},
\end{equation}
are directly tied to standard noncommutative smearings of point sources. The Gaussian arises from coherent‑state/heat‑kernel smearing and represents the fastest localised fall‑off, while the Lorentzian is a widely used alternative with a heavier tail \cite{Anacleto2020PLB, Liang2012EPL, Bhar2014EPJC}. Near the throat, these profiles are indistinguishable to quadratic order, so all near‑throat consequences, such as the NEC slope and shell width, depend only on $\delta_0=\delta(x_0)$, not on the choice of tail. The two choices, ${\delta_{G},\delta_{L}}$, thus span two natural extremes: rapid localisation with a Gaussian profile and heavy-tailed localisation with a Lorentzian one, while keeping the construction minimal and fully consistent with the noncommutative geometry–inspired matter model. Thus, whether one adopts the Gaussian or the Lorentzian profile, the quasi-de Sitter EOS constructed from $\delta(r)$ achieves the same goal, i.e. it restores the flare-out condition while driving the spacetime back to the de Sitter-like regime $p_r\simeq-\rho$ away from the throat, so that $\rho+p_r$ saturates the NEC asymptotically. The NEC violation is therefore localised in magnitude near the throat, while $\rho+p_r\to 0^{-}$ at large $x$. If one insists on strict pointwise NEC recovery beyond a finite radius, one may replace $\delta_{G, L}$ by a smooth compact support bump. We do not pursue this here since our goal is minimality and analytic/numerical simplicity. Combining \eqref{ed}, \eqref{NEC}, and \eqref{perturbation} gives
\begin{equation}\label{flare}
  b'(x_0)=\frac{1}{1+\delta_0}<1.
\end{equation}
So the flare–out condition is enforced once $\delta_0>0$. From a geometric perspective, the throat realises a spatial bounce of the areal radius, encoded in the flare–out conditions. In time-dependent cosmological models, quasi–de Sitter or Chaplygin-like fluids are often used to generate nonsingular bounces of the scale factor, where the conditions $H=0$ and $\dot{H}>0$ typically require a suitable violation of the NEC. In our static wormhole setting, the analogous role is played by the minimum of the areal radius and the flare–out condition. Equations \eqref{NEC} and \eqref{flare} show that the same combination $\rho + p_r$ that governs NEC violation also enforces this geometric bounce at the throat. A positive deformation parameter $\delta_0$ simultaneously restores the flare-out condition and localises the departure from NEC saturation near $x_0$ in the sense that $|\rho+p_r|$ decays rapidly away from the throat. Importantly, $\delta_0$ is not a free parameter because it is fixed by the throat location $x_0$ and the rescaled mass $\mu$. Using $b'(x)=\mu\,x^2 e^{-x^2/4}/\sqrt{\pi}$, one obtains
\begin{equation}\label{delta0}
  \delta_0=\frac{\sqrt{\pi}\,e^{x_0^2/4}}{\mu\,x_0^2}-1.
\end{equation}
Last but not least, notice that we choose the multiplicative perturbation \eqref{perturbation} instead of its additive counterpart because it preserves the key features of the unperturbed de Sitter fluid in a scale-free way. As soon as $\rho$ decays, the pressure vanishes at the same rate, so the EOS goes into the de Sitter regime at large radii.  Moreover, the dimensionless factor $\delta$ offers direct geometric control over the flare-out condition \eqref{flare}, thus allowing us to quantify the violation of the NEC with a single dimensionless parameter, while an additive perturbation would introduce a density scale and possible unwanted sign changes in $p_r$ outside the exotic core. Moreover, combining the field equation \eqref{eq2} with the density-pressure relation \eqref{ed} and the perturbed EOS \eqref{perturbation}, and using once again the dimensionless variable $x = r/\sqrt{\theta}$ and the rescaled mass $\mu = M/\sqrt{\theta}$, we obtain a first-order differential equation for the redshift function
\begin{equation}\label{ODEphi}
  \Phi^{'}(x)=\frac{xb^{'}(x)[1+\delta(x)]-b(x)}{2x[b(x)-x]}.
\end{equation}

\begin{table}[ht]
%\centering
\caption{Representative numerical values of the throat location $x_0$ and the corresponding evaluation of the redshift function and tangential pressure at the throat of the wormhole. The numerical values for $\Phi^{'}(x_0)$ and $p_t(x_0)$ have been evaluated by means of \eqref{Taylor} and \eqref{pt0}, respectively.}
\label{table:delta0}
\vspace*{1em}
\begin{tabular}{||c|c|c|c|c|c|c|c||}
\hline\hline
$\mu$              & $x_0$           & $\delta_0$   &$\Phi^{'}(x_0)$ & $\Phi(x_0)$ (\text{Gaussian}) & $\Phi(x_0)$ (\text{Lorentzian}) & $p_t(x_0)$ (\text{Lorentzian})\\ [0.5ex]
\hline\hline
$1.905$            & $3.0804$  & $0.0512$  &$8.9809$ &$-3.3368$  & $-3.3340$ &$-0.0249$\\
$1.950$            & $3.4659$  & $0.5246$  &$1.5355$ &$-1.9044$  & $-1.8951$ &$-0.0030$\\
$2.5$              & $4.9685$  & $12.7547$ &$1.0218$ &$-1.6003$  & $-1.5932$ &$-0.0014$\\[1ex]
\hline\hline 
\end{tabular}
\end{table}

From Table~\ref{table:delta0}, we observe that, for all considered values of the rescaled mass $\mu$, the function $\delta(x_0)$ remains strictly positive. Furthermore, at the throat, the denominator of \eqref{ODEphi} vanishes since $b(x_{0}) = x_{0}$. However, the numerator vanishes as well when  \eqref{flare} is imposed, ensuring that $\Phi^{'}(x)$ remains finite at the throat. In fact, a Taylor expansion around $x = x_0$ yields
\begin{equation}
\Phi^{'}(x)=\frac{x_0 b_1(1+\delta_0)-x_0}{2x_0(b_1-1)}\frac{1}{x-x_0}
+\frac{b_1[4+b_2 x_0(1+\delta_0)]-2(1+b_1^2)-b_2 x_0(1+2\delta_0)}{4x_0^2(b_1-1)^2}+\mathcal{O}(x-x_0).
\end{equation}
and since \eqref{flare} implies that $b_1(1+\delta_0)=1$, we end up with
\begin{equation}\label{Taylor}
\Phi^{'}(x)=-\frac{1}{2x_0}-b_2\left[1+\frac{1}{2}\left(\delta_0+\frac{1}{\delta_0}\right)\right]+\mathcal{O}(x-x_0),
\end{equation}
where $b_1$ and $b_2$ has been defined in Section~\ref{Sec2}. Regarding asymptotic flatness, observe that as $x\to\infty$, we have $b(x)\to 1$, $b^{'}\to 0$, $xb^{'}\to 0$, and $\delta(x)\to 0$. Hence, from the ODE \eqref{ODEphi}, we can immediately read out the leading fall-off behaviour
\begin{equation}
\Phi^{'}(x)=\frac{\mu}{x^2}+\mathcal{O}\left(\frac{1}{x^3}\right).
\end{equation}
Integrating once,
\begin{equation}
    \Phi(x)=\Phi_{\infty}-\frac{\mu}{x}+\mathcal{O}\left(\frac{1}{x^2}\right)\,.
\end{equation}
So the redshift function tends to a finite constant $\Phi_\infty$. That constant is physically harmless because a transformation $t\mapsto e^{-\Phi_\infty}t$ rescales the time coordinate and sends $\Phi_\infty\to 0$. This is implemented numerically by choosing the throat value $\Phi(x_0)=-\Phi_\infty$ so that the integrated solution automatically satisfies $\Phi(\infty)=0$, ensuring an asymptotically flat metric without further adjustment. These properties ensure that the quasi-de Sitter fluid with perturbation $\delta(x)$ generates a redshift function that is both regular at the throat and asymptotically flat, satisfying the necessary conditions for a traversable wormhole. Concerning the tangential pressure, if we replace \eqref{ODEphi} into the rescaled version of \eqref{pt}, expand around $x=x_0$ and make use of \eqref{flare}, we find that it exhibits a finite behaviour at the throat
\begin{equation}\label{pt0}
p_t(x)=-\frac{x_0^2(b_2^2+b_3)\delta_0^3-\left[2+4x_0+x_0^2(2-2b_3-3b_2^2)\right]\delta_0^2+x_0^2\delta_0(3b_2^2+b_3)+b_2^2 x_0^2}
{64\pi\delta_0 x_0^3(1+\delta_0)}+\mathcal{O}(x-x_0),    
\end{equation}
where $b_3=b^{'''}(x_0)$.

\begin{figure}[!ht]
\centering
    \includegraphics[width=0.4\textwidth]{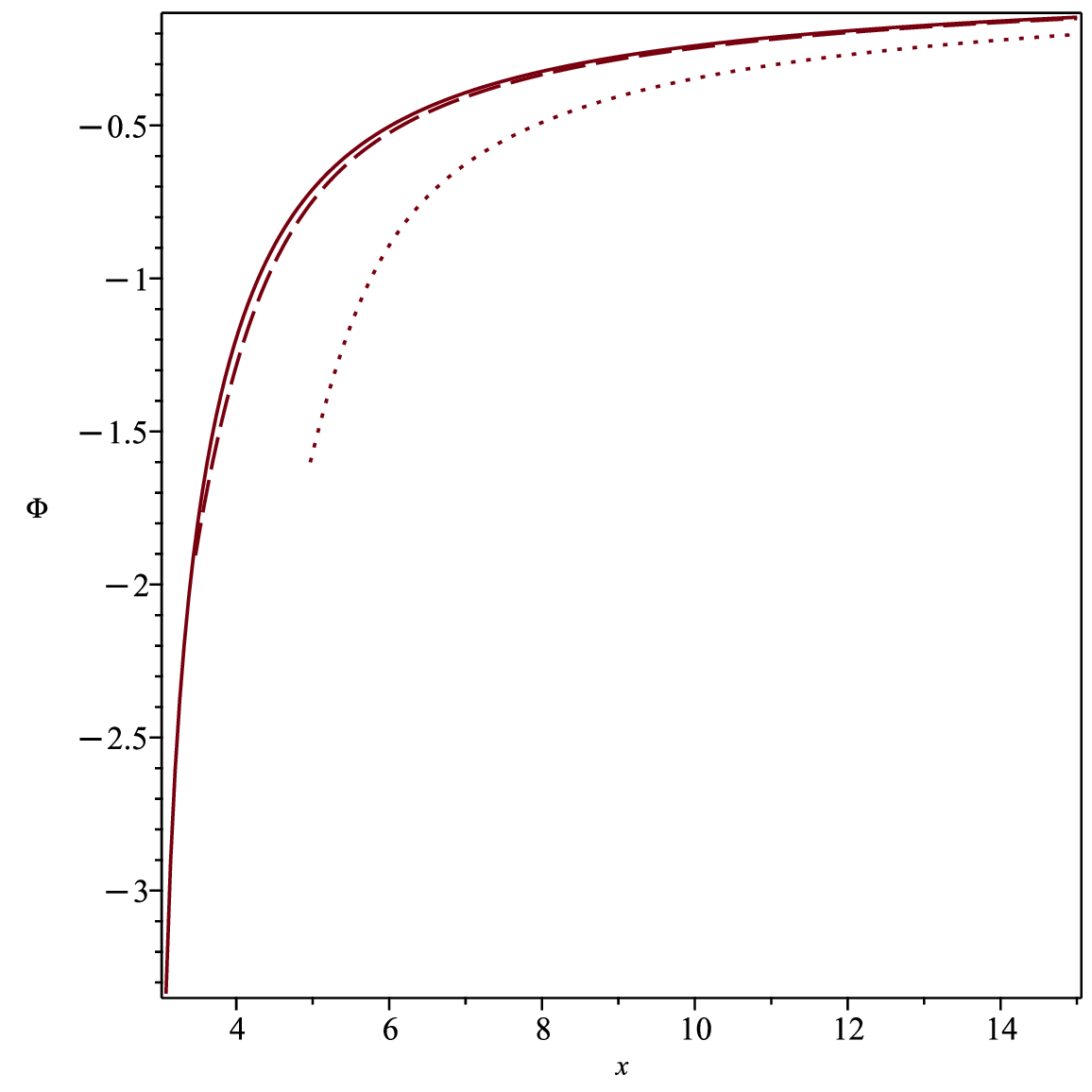}
    \includegraphics[width=0.4\textwidth]{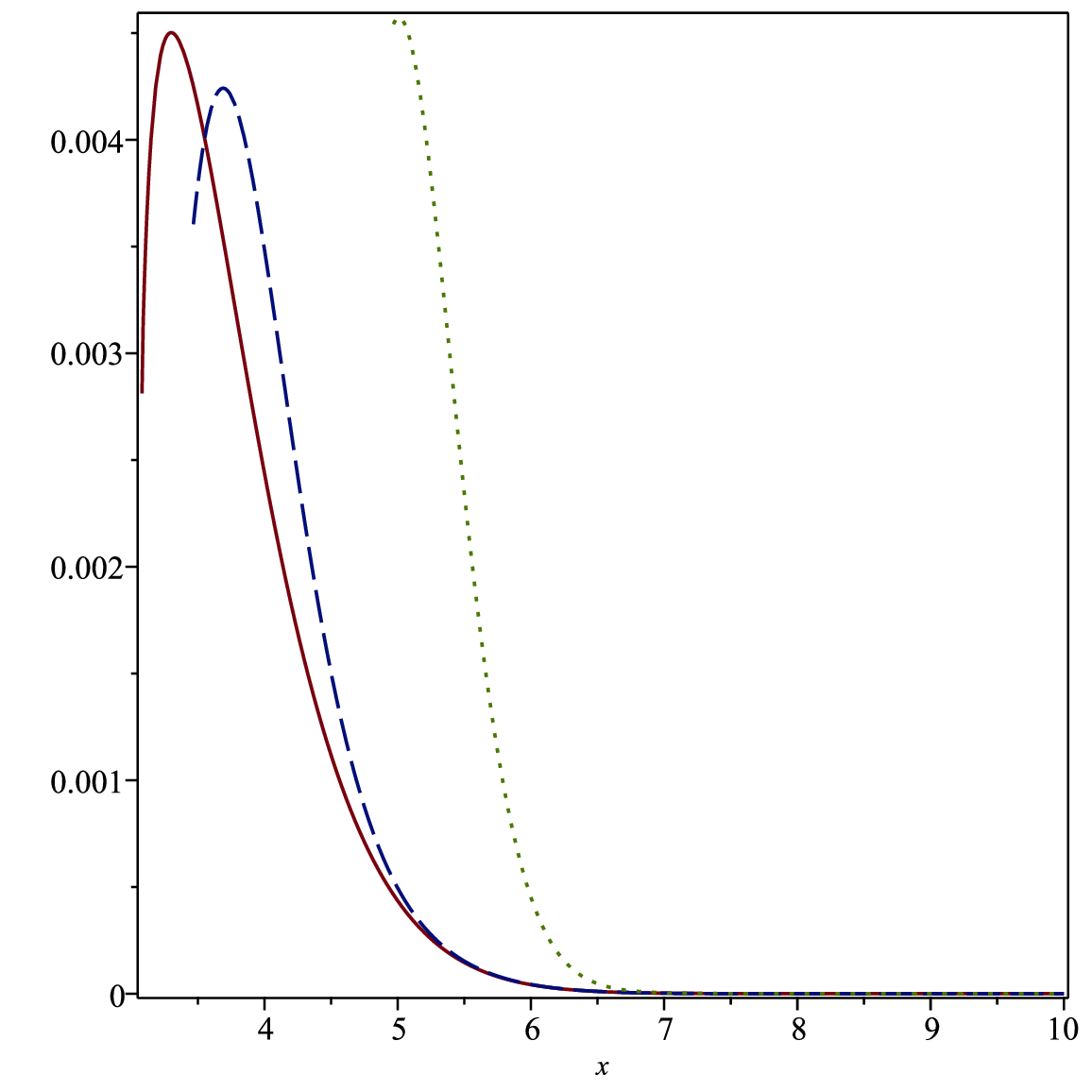}
\caption{\label{PhiP1} {\bf{Left panel}}: Redshift profile $\Phi$ as a function of the dimensionless radius $x$ for three wormhole solutions obeying a Gaussian quasi-de Sitter EOS. Solid line ($\mu=1.905$), dashed line ($\mu=1.95$), and dotted line ($\mu=2.5$). The corresponding throat values $\Phi(x_0)$ are listed in Table~\ref{table:delta0}. {\bf{Right panel}}: Plot of the dimensionless tangential pressure $p_t$ for the same three masses.}
\end{figure}

\subsection{Model A: quasi–de Sitter EOS with Gaussian bump}

We begin by modelling the departure from the exact de Sitter EOS with a Gaussian bump centred at the throat, as given in the first relation of \eqref{deltaGL}. Figure~\ref{PhiP1} displays the corresponding redshift function obtained by integrating \eqref{ODEphi} numerically using a Runge–Kutta–Fehlberg (RKF4/5) scheme in \textsc{Maple}. The throat value is fixed to $\Phi(x_0)=-\Phi_\infty$, which guarantees asymptotic flatness and ensures that the redshift function vanishes at infinity. The left panel shows that increasing $\mu$ mainly affects the immediate vicinity of the throat. The three redshift profiles deviate noticeably close to $x_0$, but beyond a few throat radii they rapidly converge and become practically indistinguishable. The right panel of Fig.~\ref{PhiP1} complements this by plotting the tangential pressure $p_t$ for the same choices of the rescaled mass parameter. In all cases, $p_t$ develops a single positive peak just outside the throat, demonstrating that the matter distribution is strongly anisotropic only in a narrow neighbourhood of $x_0$. As $\mu$ increases, the peak becomes taller while its width decreases, indicating that the anisotropy is increasingly localised. For $x\gtrsim 7$, all curves merge into a common tail. Moreover, none of the profiles crosses the axis, and hence, the tangential pressure remains non-negative everywhere. This confirms that the anisotropic stresses are confined to a small region close to the throat and decay rapidly away from it. In particular, since $\delta(x)$ is localised around $x_0$ and vanishes at large radii, the combination $\rho + p_r = -\rho\delta(x)$ is negative only in a thin neighbourhood of the throat and tends to zero from below as $x\to\infty$, so the NEC violation is limited to a narrow shell around $x_0$.
\begin{figure}[!ht]
\centering
    \includegraphics[width=0.4\textwidth]{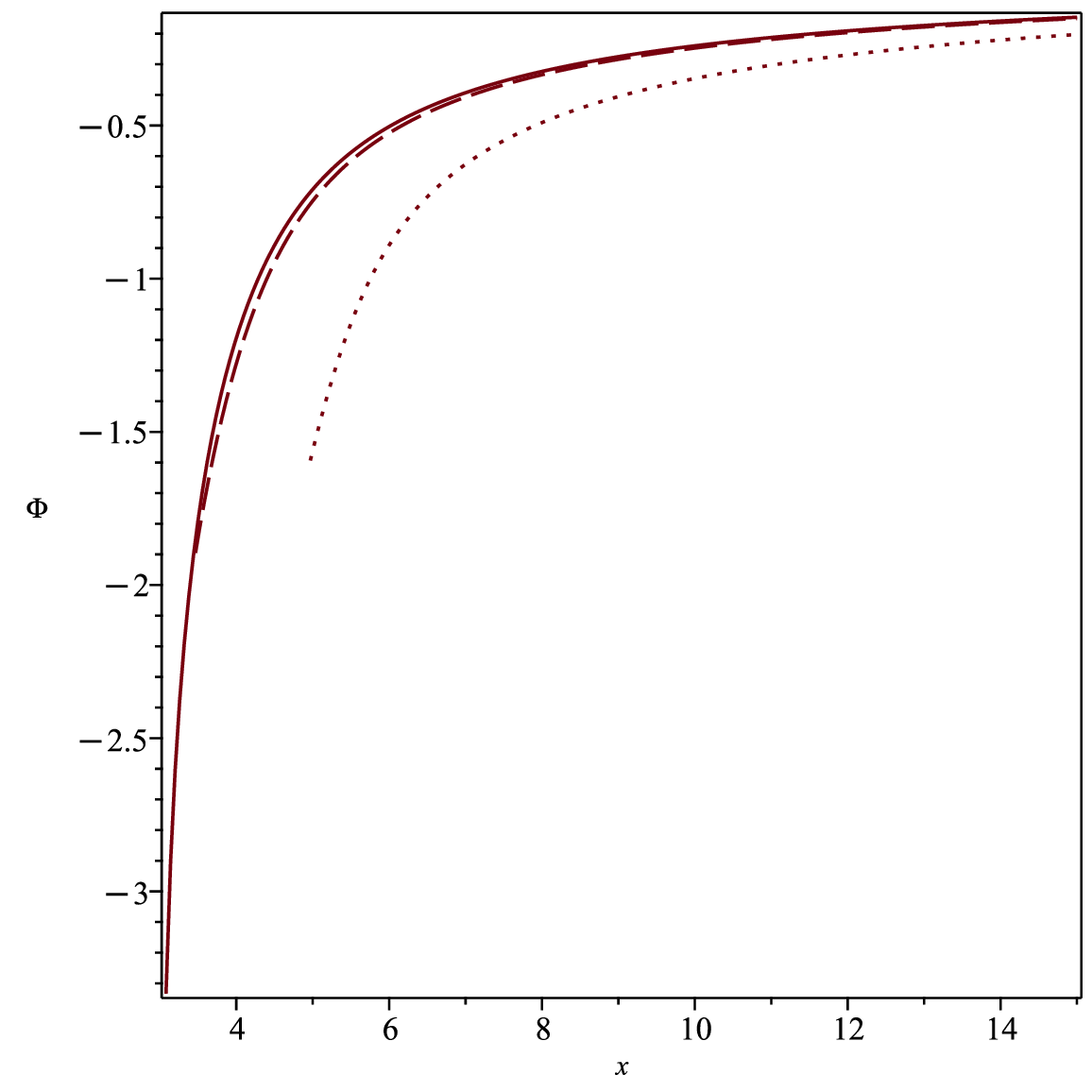}
    \includegraphics[width=0.4\textwidth]{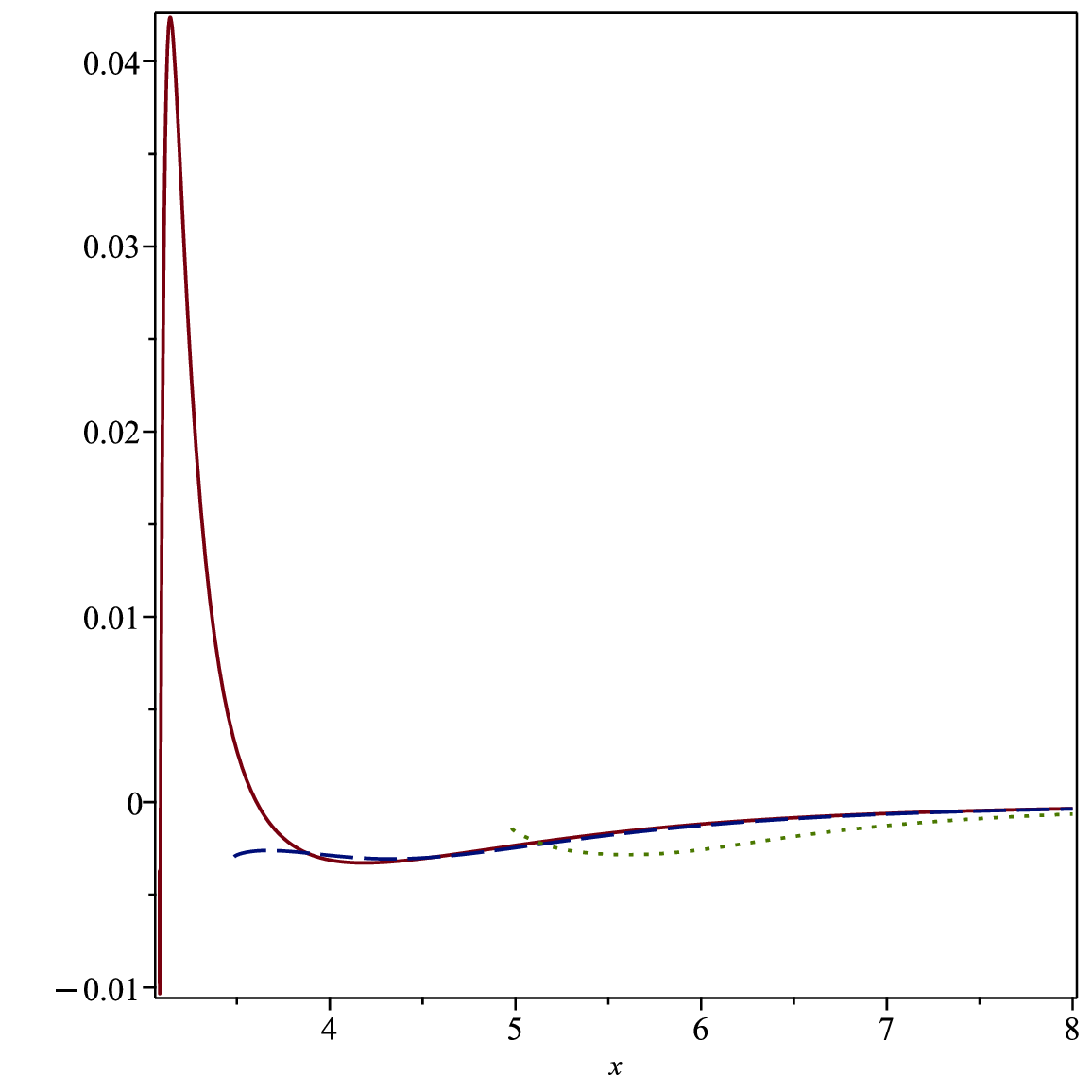}
\caption{\label{PhiP3} {\bf{Left panel}}: Redshift profile $\Phi$ as a function of the dimensionless radius $x$ for three wormhole solutions obeying a Lorentzian quasi-de Sitter EOS. Solid line ($\mu=1.905$), dashed line ($\mu=1.95$), and dotted line ($\mu=2.5$). The corresponding throat values $\Phi(x_0)$ are listed in Table~\ref{table:delta0}. {\bf{Right panel}}: Plot of the dimensionless tangential pressure $p_t$ for the same three rescaled masses. The numerical values at the throat are displayed in Table~\ref{table:delta0}.}
\end{figure}

\subsection{Model B: quasi–de Sitter EOS with Lorentzian bump}

Let us consider now a deviation from the exact de-Sitter EOS in terms of a Lorentzian bump centred on the throat, as given by the second formula in \eqref{deltaGL}. Figure~\ref{PhiP3} displays the redshift and tangential pressure profiles obtained in this scenario.  Since the Gaussian and Lorentzian perturbations $\delta_G$ and $\delta_L$ coincide up to order $\mathcal{O}(x-x_0)^3$ and first differ only at $\mathcal{O}(x-x_0)^4$, the corresponding redshift functions, once integrated with essentially the same initial data, can differ at most by $\mathcal{O}(x-x_0)^3$ within the thin region where the bumps act. As a result, the two redshift profiles are practically indistinguishable. However, pronounced differences emerge in the tangential pressure. With a Lorentzian quasi–de Sitter EOS, $p_t$ behaves very differently in contrast with the Gaussian case, where $p_t$ remained everywhere non-negative. For $\mu=1.905$, the profile is negative right outside the throat, then rises and crosses to positive, reaches a positive maximum, drops back below zero to a negative minimum, and finally tends to zero from below at large values of $x$. For $\mu=1.95$, the pressure is negative everywhere. It lifts slightly after the throat to a small negative peak, then falls to a negative minimum before relaxing asymptotically. For $\mu=2.5$, the pressure is again negative everywhere. It decreases from the throat to a single negative minimum and then returns toward zero. Thus, increasing $\mu$ suppresses the brief positive overshoot seen in the nearly extremal case $\mu=1.905$ and keeps $p_t<0$ throughout, while the far-field approach remains universal. Compared with Model A, the Lorentzian bump has a heavier tail, so the region where $\rho + p_r < 0$ is slightly more extended, but the magnitude of $\rho+p_r$ remains slightly more extended due to the heavier tail of $\delta_L$, but $\rho+p_r$ still tends to $0^{-}$ asymptotically as $\delta_L(x)\to 0$.

\subsection{Model C: Chaplygin-like EOS}

In addition to Gaussian or Lorentzian perturbations, we also consider a Chaplygin-inspired EOS of the form
\begin{equation}\label{chap}
p_r(x)=-\rho(x)\left[1+\left(\frac{A}{\rho^{\alpha+1}(x)}-1\right)e^{-(x-x_0)^2}\right],\quad \alpha>0,\quad
A=\rho^{\alpha+1}(x_0)\left(1-\frac{b^{'}(x_0)-1}{8\pi x_0^2\rho(x_0)}\right).
\end{equation}
At the throat $x = x_0$, the exponential factor in \eqref{chap} equals unity, and the EOS reduces to a generalised Chaplygin form, $p_r(x_0)=-A/\rho^\alpha(x_0)$. On the other hand, for $x\gg x_0$, the exponential term tends to zero and the bracket in \eqref{chap} approaches unity, so that $p_r(x) \to -\rho(x)$. In this sense, the Chaplygin-like EOS interpolates between a local Chaplygin behaviour near the throat
and a de Sitter-like relation $p_r = -\rho$ in the asymptotic region. Notice that $A$ has been chosen so that the NEC condition \eqref{NEC} is automatically satisfied. Furthermore, the radial pressure vanishes as $x\to+\infty$ provided that $\alpha<4$. This construction is motivated by the generalised Chaplygin EOS $p=-A/\rho^{\alpha}$ \cite{Bento2002PRD, Eiroa2009PRD}, widely studied in cosmology as a unification model for dark matter and dark energy and also used in wormhole physics \cite{Jamil2009EPJC, Lobo2005PRD, Kuhfittig2009GRG, Rahaman2009APPB}. Here, the exponential factor localises the non-linear Chaplygin behaviour to a neighbourhood of the throat, ensuring that the asymptotic region remains de Sitter-like. The free parameter $\alpha$ controls the strength of the nonlinearity, while the adiabatic sound speed $c_s^2=\partial p_r/\partial\rho$ provides a natural criterion to restrict its range by demanding subluminal propagation. Even though, in an anisotropic fluid we can, in principle, define two sound speeds $c^2_{s,r}=\partial p_r/\partial\rho$ and $c^2_{t,r}=\partial p_t/\partial\rho$, in our set-up, only the radial pressure is prescribed by an EOS. Hence, $p_t$ is not an independent constitutive law because it is determined indirectly by the metric via $\Phi$, $b$, once $p_r(\rho,r)$ is chosen. That is why the physically meaningful diagnostic for causality of the EOS we have introduced is the radial one, namely $c_s=c_{s,r}$. More specifically, for our localised, Chaplygin-inspired EOS, we find
\begin{equation}\label{sound}
  c^2_s(x)= -1+e^{-(x-x_0)^2}+\frac{A\alpha}{\rho^{\alpha+1}(x)}e^{-(x-x_0)^2}.
\end{equation}
Away from the throat, as $x \to +\infty$, the exponential factor in \eqref{sound} tends to zero, the EOS approaches the de Sitter relation $p_r = -\rho$, and $c_s^2(x) \to -1$. This reflects the vacuum‑like character of the de Sitter–type sector, where a hydrodynamic sound speed is not physically meaningful. The behaviour of $c_s^2(x)$ for different values of the rescaled mass parameter is illustrated in the plots. The plots show that if $\alpha$ is chosen too large, $c_s^2(x)$ can exceed unity at some radius away from the throat, leading to a violation of the subluminal propagation condition. For $\mu = 1.905$ a safe choice avoiding such causality violations is $0 < \alpha \leq 0.35$, while for $\mu = 1.95$ we must choose $0 < \alpha \leq 0.23$. Finally, for $\mu = 2.5$ the bound tightens to $0 < \alpha \leq 0.016$. Thus, by restricting ourselves to these ranges of $\alpha$, the causality condition $0 < c_s^2(x) \leq 1$ is satisfied throughout the region where $c_s^2(x) \geq 0$ and the fluid interpretation of the Chaplygin-like EOS is applicable.

\begin{figure}[!ht]
\centering
    \includegraphics[width=0.32\textwidth]{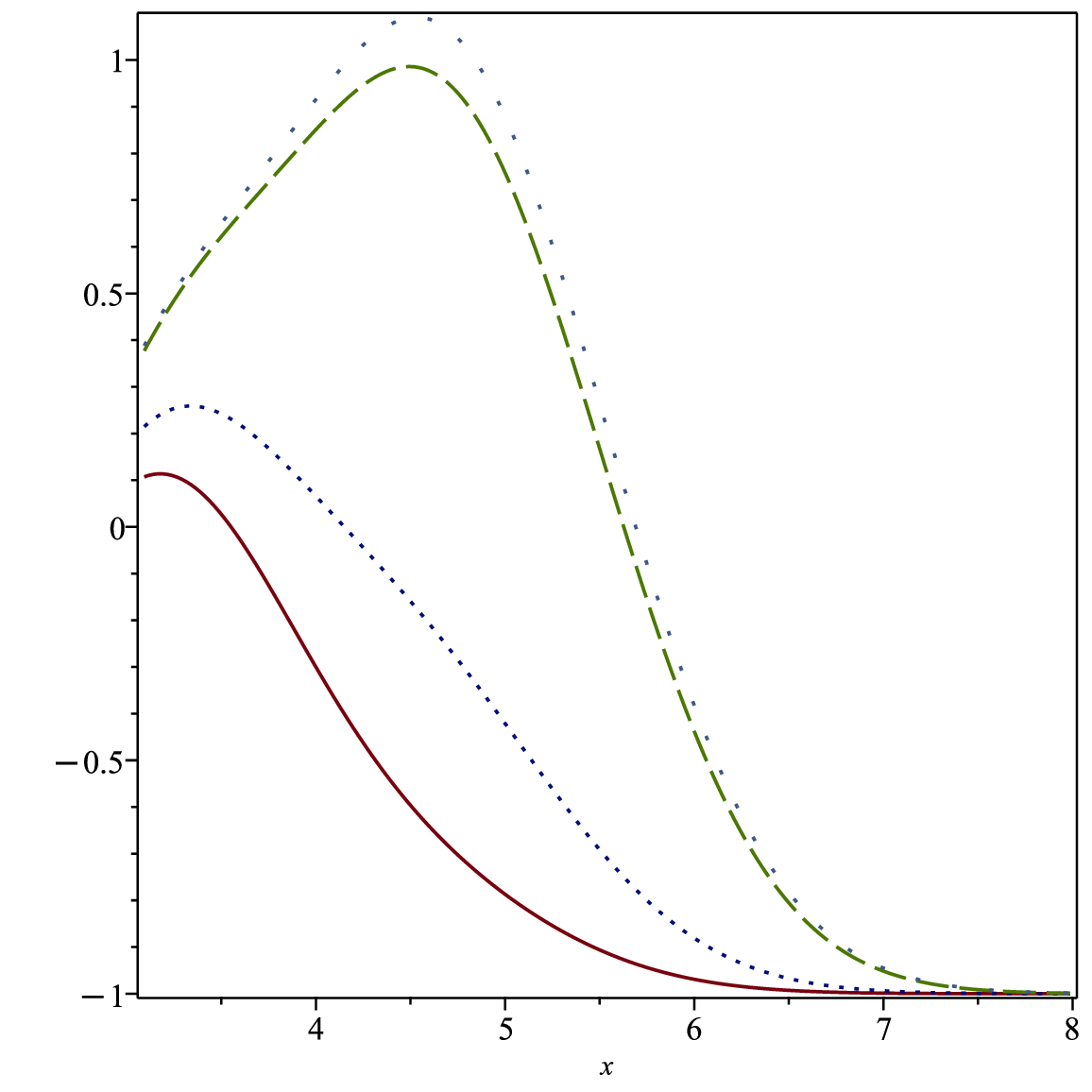}
    \includegraphics[width=0.32\textwidth]{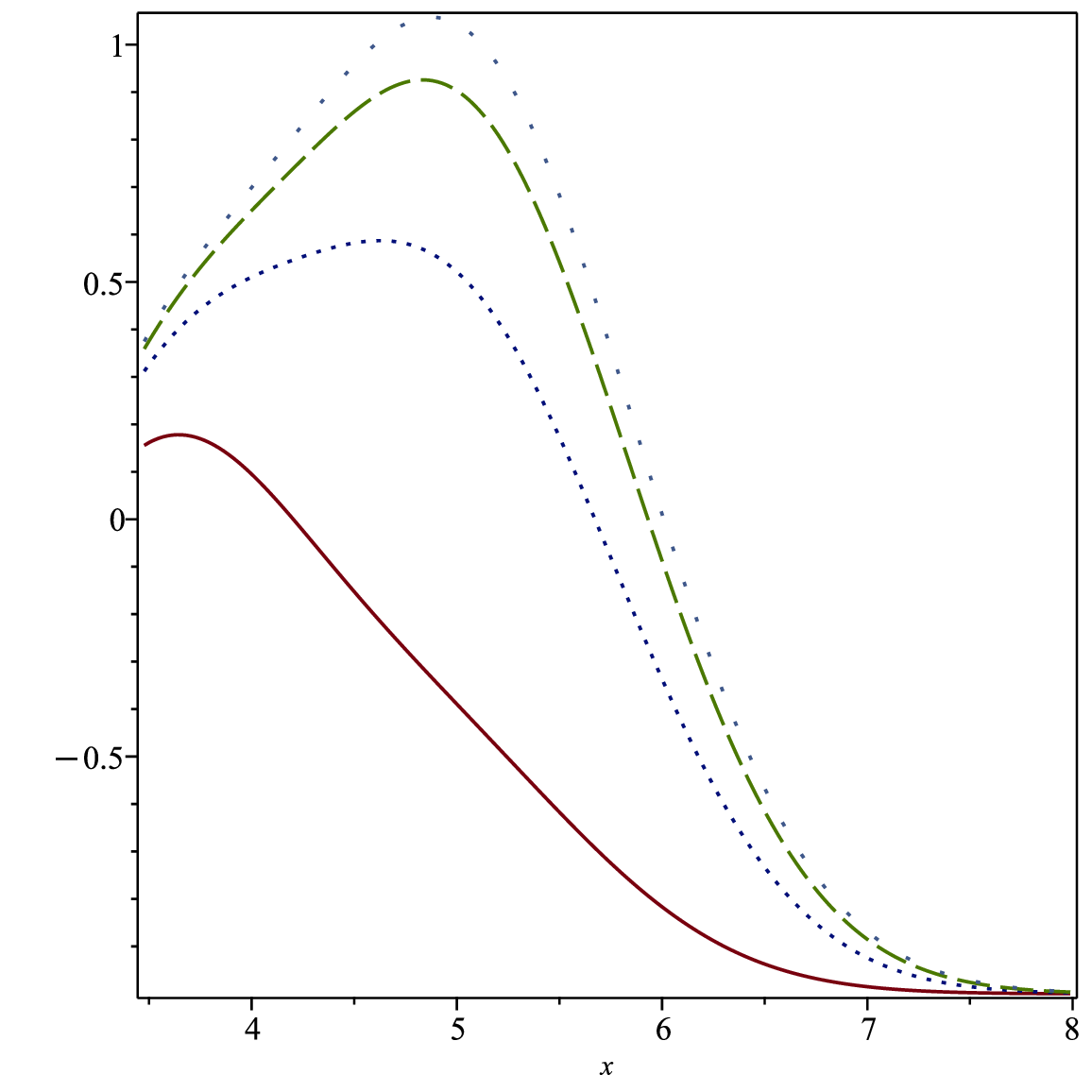}
    \includegraphics[width=0.32\textwidth]{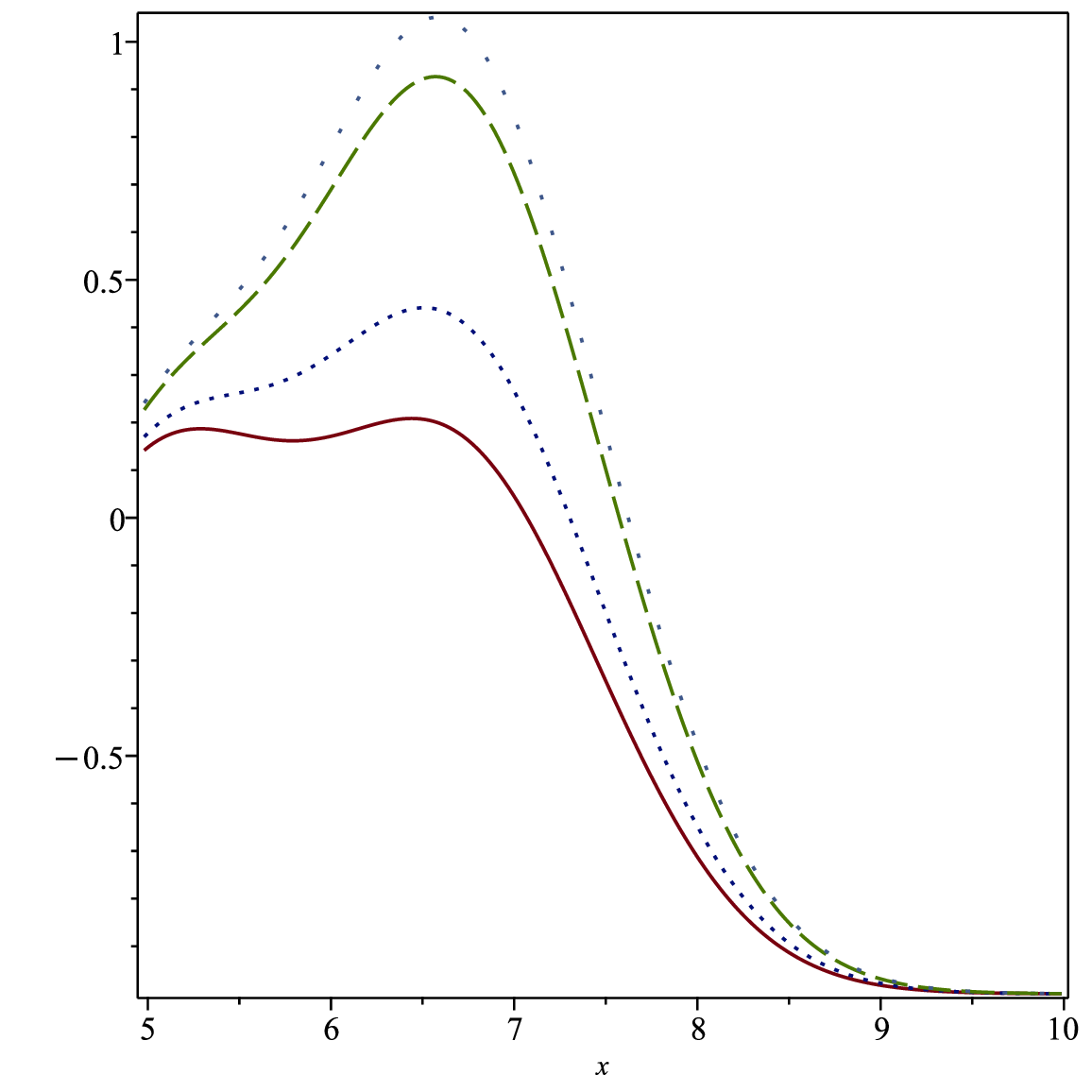}
\caption{\label{cs2} {\bf{Left panel}}: Profile of $c_s^2(x)$ as given by \eqref{sound} for $\mu=1.905$ as a function of the dimensionless radius $x$ for different values of the parameter $\alpha$. Solid line ($\alpha=0.1$), dotted line ($\alpha=0.2$), dashed line ($\alpha=0.35$), and space-dotted line ($\alpha=0.36$) {\bf{Central panel}}: Plot of $c_s^2(x)$ for $\mu=1.95$ and $\alpha=0.1$ (solid line), $\alpha=0.2$ (dotted line), $\alpha=0.23$ (dashed line), and $\alpha=0.24$ (space-dotted line). {\bf{Right panel}}: Plot of $c_s^2(x)$ for $\mu=2.5$ and $\alpha=0.01$ (solid line), $\alpha=0.012$ (dotted line), $\alpha=0.016$ (dashed line), and $\alpha=0.017$ (space-dotted line).}
\end{figure}

\begin{figure}[!ht]
\centering
    \includegraphics[width=0.4\textwidth]{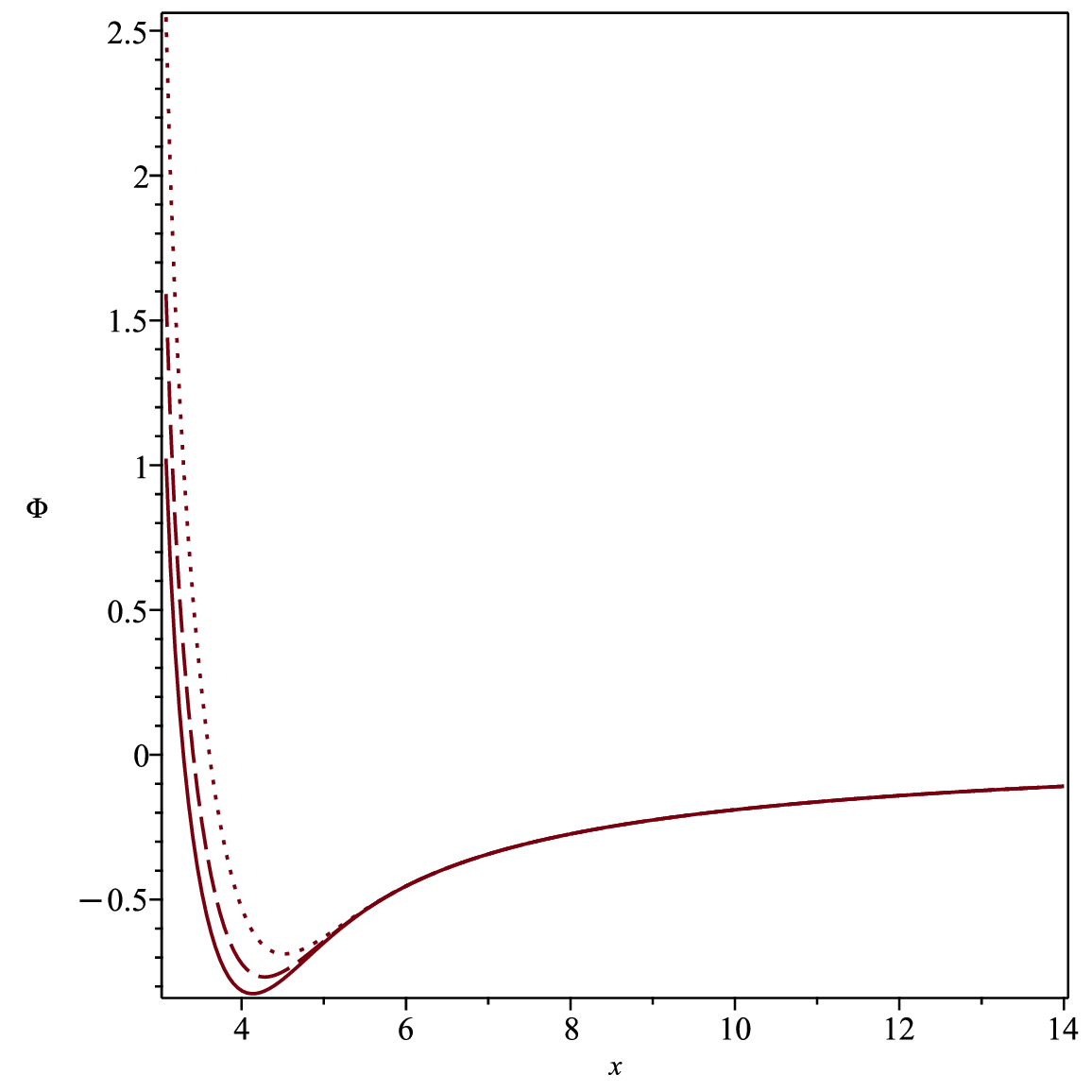}
    \includegraphics[width=0.4\textwidth]{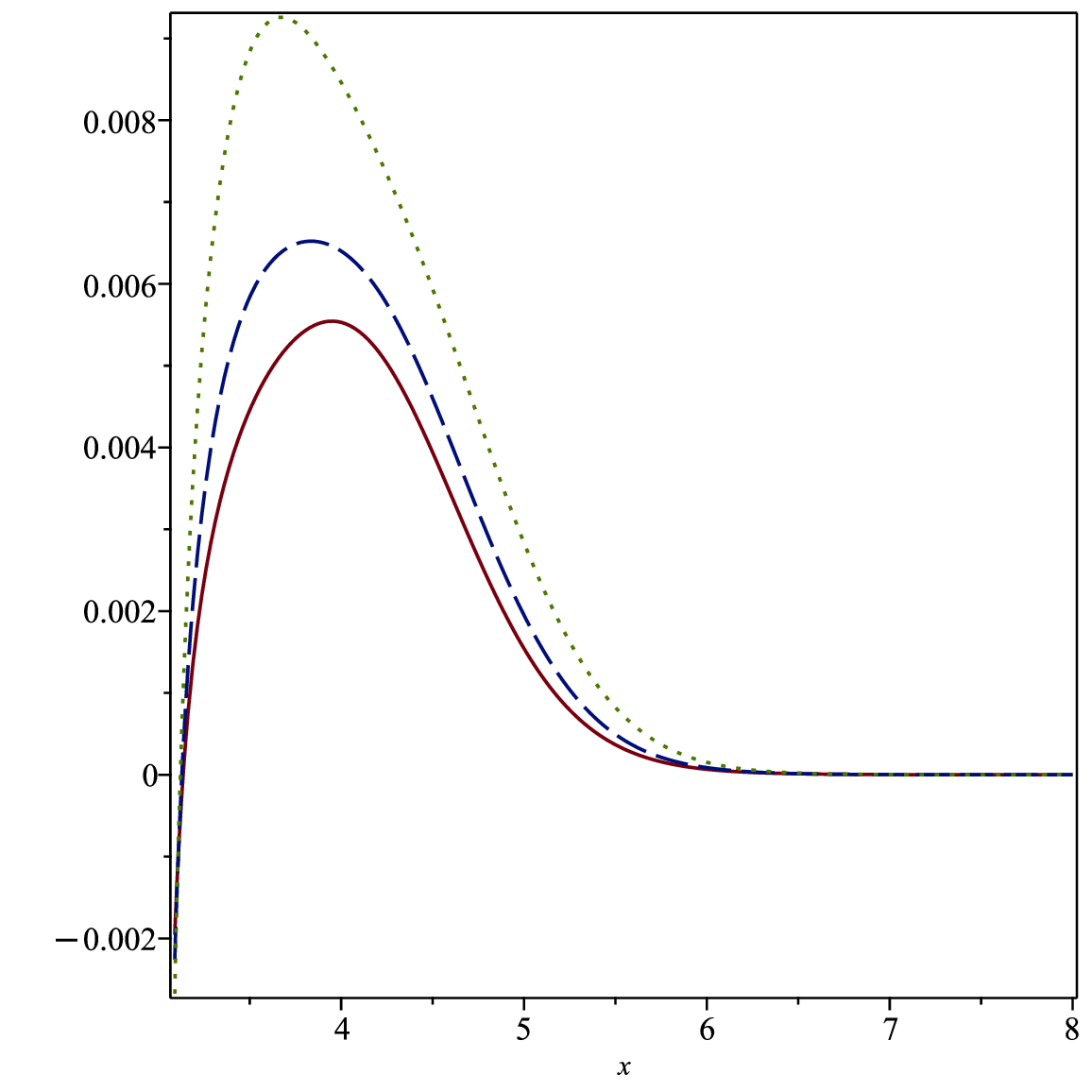}
\caption{\label{PhiP2} {\bf{Left panel}}: Redshift profile $\Phi$ for $\mu=1.905$ as a function of the dimensionless radius $x$ for three wormhole solutions obeying the Chaplygin-like EOS \eqref{chap}. Solid line ($\alpha=0.1$), dashed line ($\alpha=0.2$), and dotted line ($\alpha=0.35$). The corresponding throat values are $\Phi(x_0)=1.0223$, $1.5909$, and $2.5468$. {\bf{Right panel}}: Plot of the dimensionless tangential pressure $p_t$ for the same values of $\alpha$ and $\mu$.}
\end{figure}

\begin{figure}[!ht]
\centering
    \includegraphics[width=0.4\textwidth]{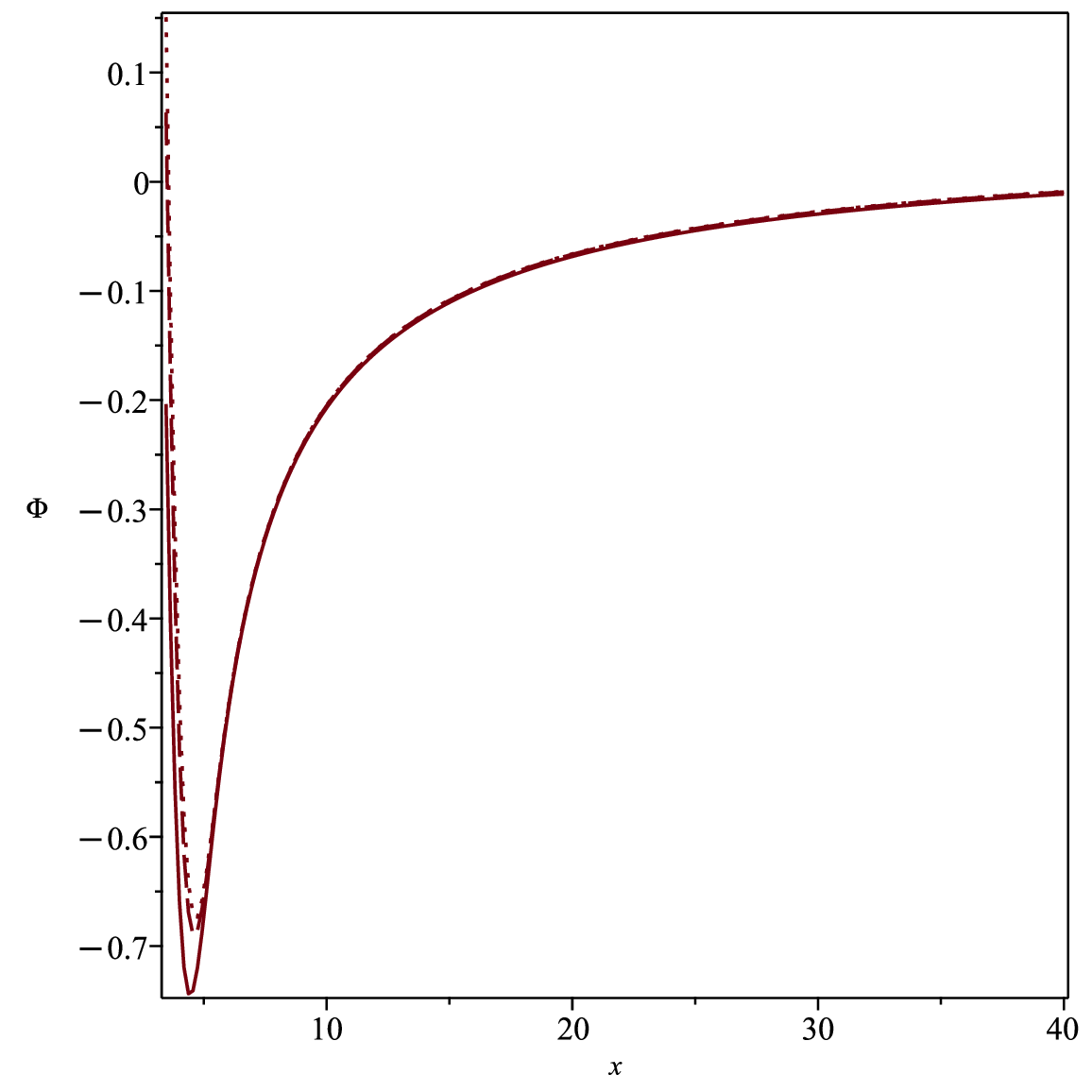}
    \includegraphics[width=0.4\textwidth]{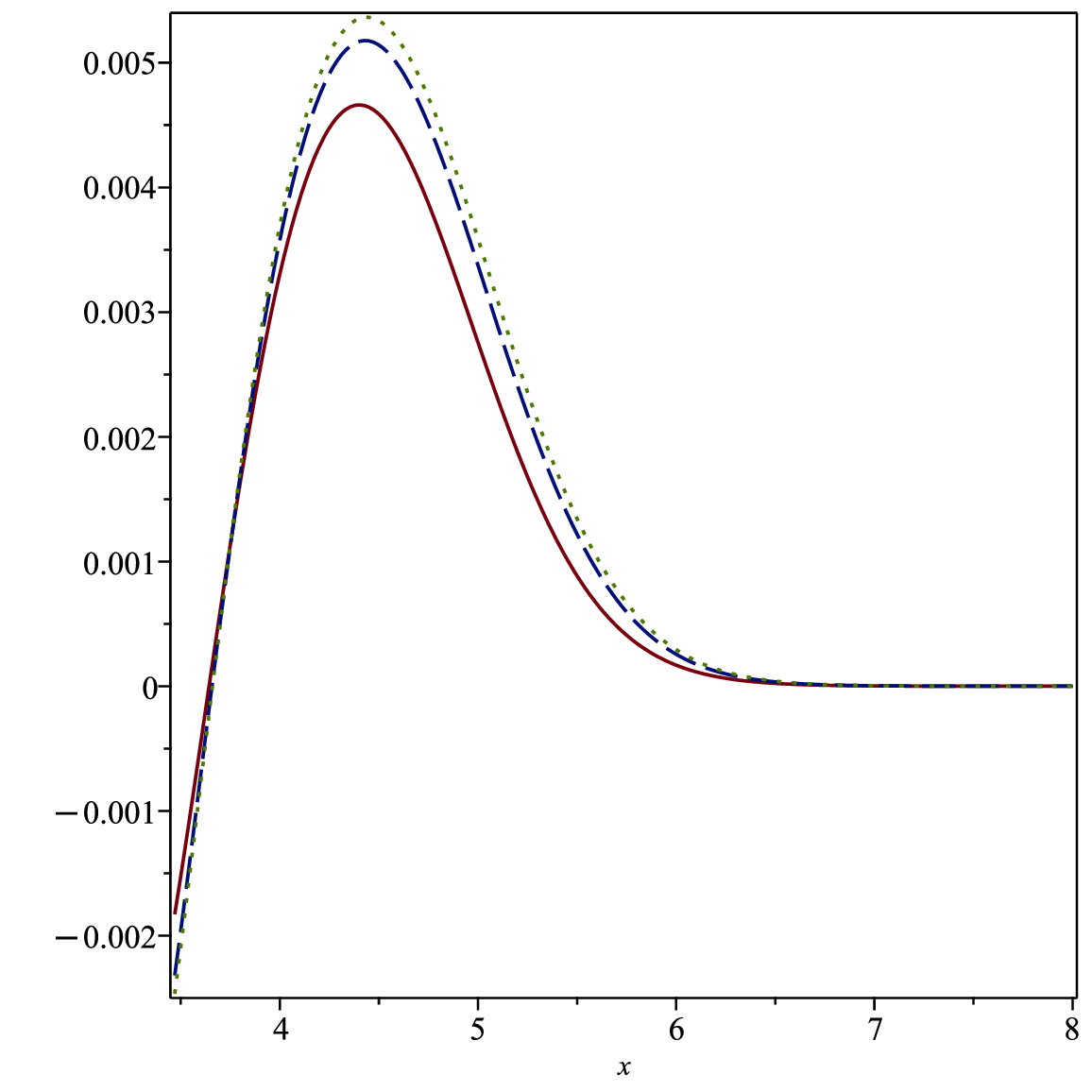}
\caption{\label{PhiP3C} {\bf{Left panel}}: Redshift profile $\Phi$ for $\mu=1.95$ as a function of the dimensionless radius $x$ for three wormhole solutions obeying the Chaplygin-like EOS \eqref{chap}. Solid line ($\alpha=0.1$), dashed line ($\alpha=0.2$), and dotted line ($\alpha=0.23$). The corresponding throat values are $\Phi(x_0)=-0.2036$, $0.0634$, and $0.1507$. {\bf{Right panel}}: Plot of the dimensionless tangential pressure $p_t$ for the same values of $\alpha$ and $\mu$.}
\end{figure}

\begin{figure}[!ht]
\centering
    \includegraphics[width=0.4\textwidth]{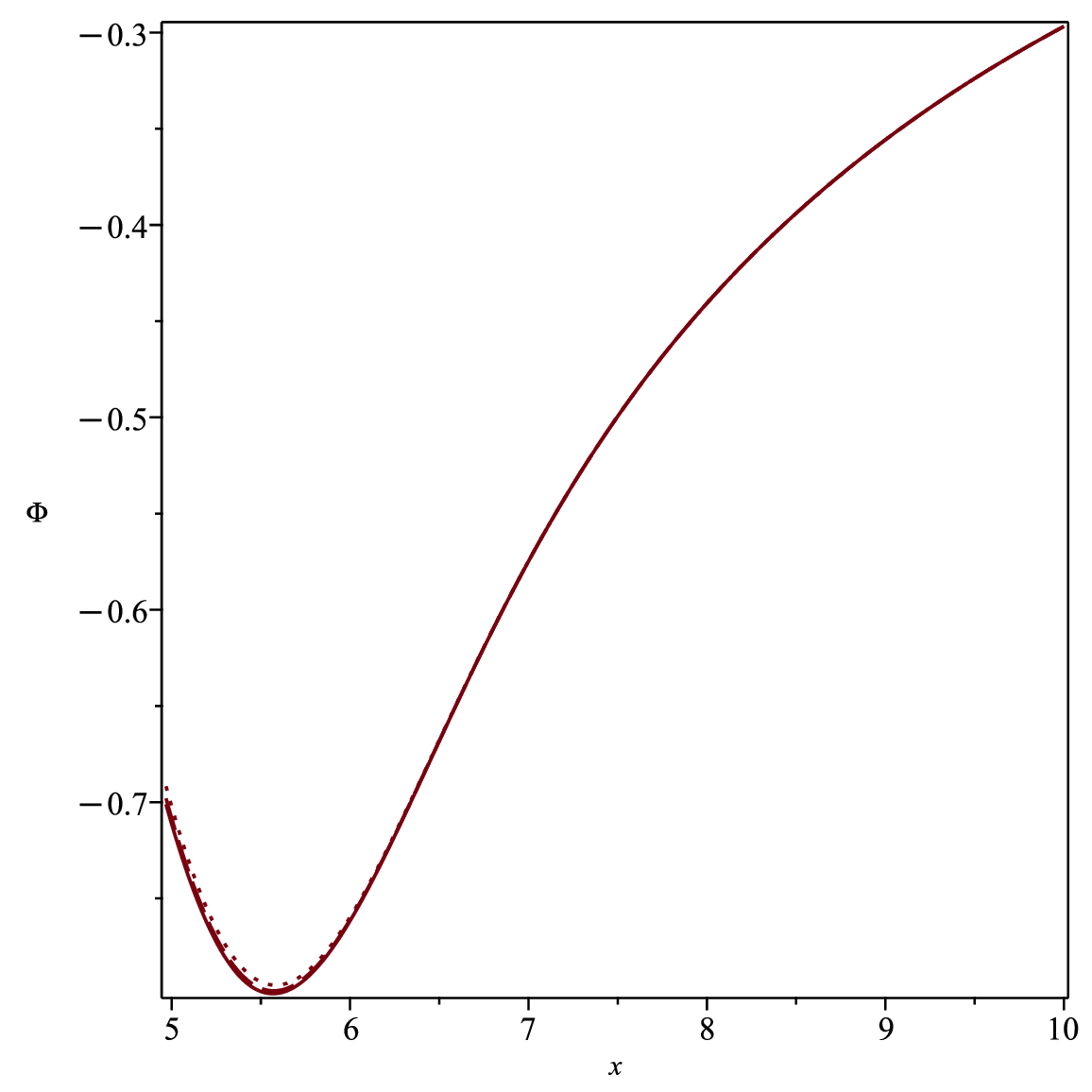}
    \includegraphics[width=0.4\textwidth]{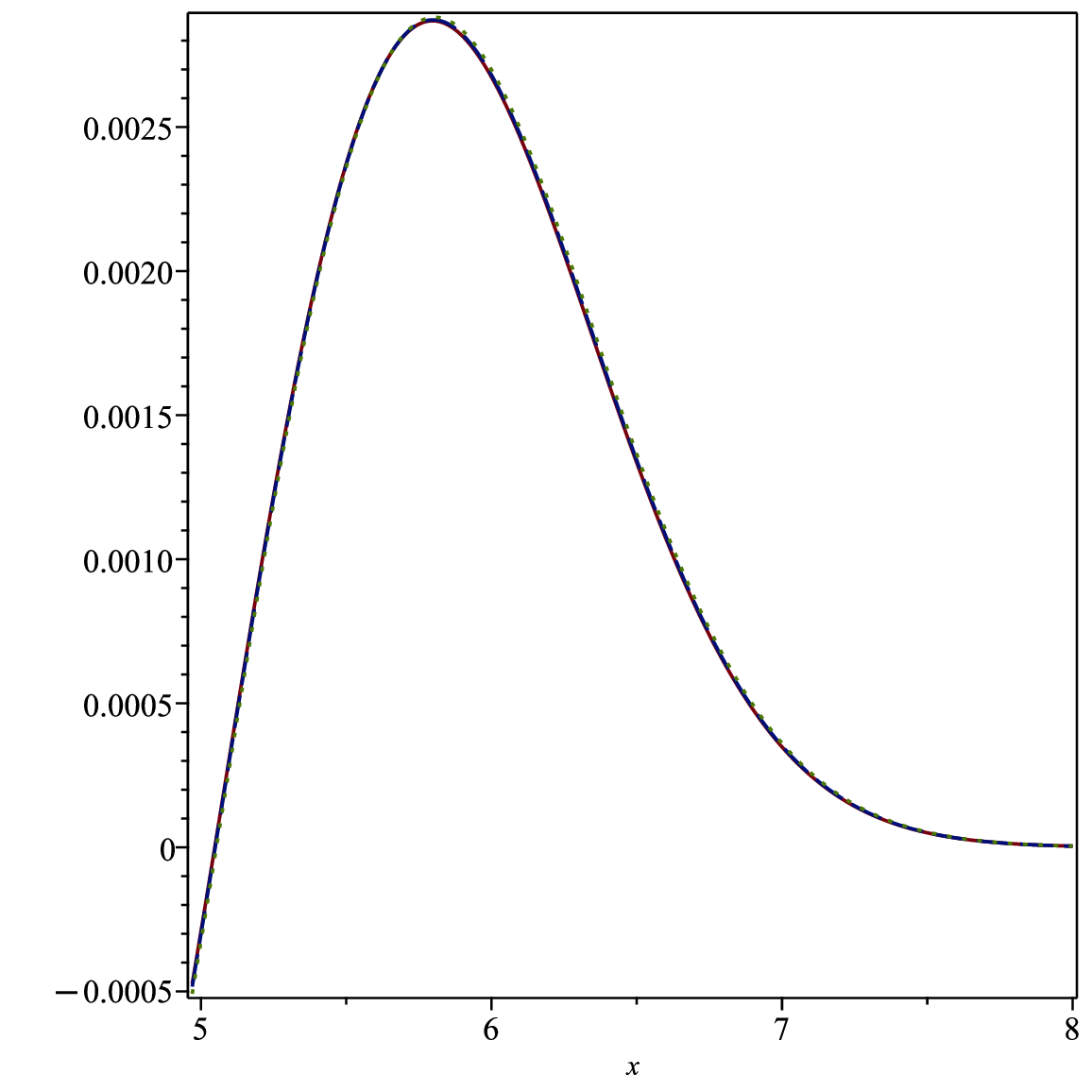}
\caption{\label{PhiP4} {\bf{Left panel}}: Redshift profile $\Phi$ for $\mu=2.5$ as a function of the dimensionless radius $x$ for three wormhole solutions obeying the Chaplygin-like EOS \eqref{chap}. Solid line ($\alpha=0.01$), dashed line ($\alpha=0.012$), and dotted line ($\alpha=0.016$). The corresponding throat values are $\Phi(x_0)=-0.7012$, $-0.6981$, and $-0.6918$. {\bf{Right panel}}: Plot of the dimensionless tangential pressure $p_t$ for the same values of $\alpha$ and $\mu$.}
\end{figure}

The Chaplygin-like EOS produces a qualitatively distinct behaviour of the redshift and tangential pressure compared with the Gaussian or Lorentzian perturbations. In the case of nearly extremal wormholes, the redshift $\Phi(x)$ starts positive at the throat, indicating a local blueshift region where proper time runs faster than at infinity, but then decreases, becomes negative, reaches a shallow minimum, and finally approaches zero as $x\to\infty$ (see Figure~\ref{PhiP2}).  As the nonlinearity parameter $\alpha$ increases toward its upper bound, the negative part of the profile becomes progressively less pronounced, and the minimum flattens, showing that stronger Chaplygin nonlinearity mitigates the redshift well. The tangential pressure $p_t(x)$ is negative in a narrow region adjacent to the throat and then turns positive, developing a single maximum just outside $x_0$.  The height of this maximum and the width of the profile grow with $\alpha$, thus implying that the anisotropic stresses become more spread as the EOS departs further from the de Sitter limit.  All $p_t$ profiles vanish asymptotically, with those corresponding to larger $\alpha$ decaying slightly more slowly.  Overall, the Chaplygin-like EOS yields redshift and pressure distributions that remain regular and asymptotically flat, but exhibit enhanced, increasingly spread anisotropies as the parameter $\alpha$ increases.

As the mass parameter increases (see Figure~\ref{PhiP3C} for $\mu=1.95$), the redshift at the throat changes character. For $\alpha=0.1$ one finds $\Phi(x_0)<0$, while for larger values of $\alpha$ it becomes positive, signalling the onset of a local blueshift region.  For all three $\alpha$ considered, $\Phi(x)$ develops a negative minimum just outside the throat before approaching zero as $x\to +\infty$. Such a minimum becomes progressively shallower as $\alpha$ increases, indicating that the Chaplygin nonlinearity weakens the redshift well. The tangential pressures display the same qualitative structure as in the nearly extremal case $\mu=1.905$. More precisely, $p_t$ is negative in a narrow interval adjacent to the throat and then turns positive, reaching a single maximum before relaxing asymptotically to zero.  For $\mu=1.95$, the interval where $p_t<0$ is slightly broader, indicating that the region of tangential tension extends farther from the throat as the configuration departs from extremality, while the overall amplitude of the anisotropy remains modest.

For the higher mass parameter $\mu = 2.5$ and the corresponding values of $\alpha$ shown in Figure~\ref{PhiP4}, the redshift remains negative at the throat for all cases considered. Each profile develops a shallow negative minimum of comparable depth to those found for lower values of $\mu$. However, because the admissible range of $\alpha$ is extremely narrow for this mass, the redshift profiles show very little sensitivity to variations in $\alpha$. This indicates that the Chaplygin-like nonlinearity becomes progressively less influential as $\mu$ increases. The tangential pressures exhibit the same qualitative structure as in the previous cases. $p_t$ is negative in a narrow region near the throat, becomes positive, and develops a single maximum. Here too, the $p_t$ profiles vary only mildly with $\alpha$, again reflecting the small parameter window allowed by causality. Compared with the nearly extremal configuration, the height of the $p_t$ peak is reduced, showing that the anisotropic stresses weaken as $\mu$ increases even though their overall shape remains unchanged. All profiles converge smoothly to zero at large $x$, confirming that the anisotropy remains confined to a
limited neighbourhood of the throat.

Taken together, the Chaplygin-like EOS exhibits a behaviour distinct from that of the Gaussian or Lorentzian perturbations.  While all three constructions confine the anisotropy to a thin neighbourhood of the throat, the Chaplygin model introduces a nonlinear coupling between pressure and density, thereby modifying the redshift potential. The resulting redshift functions may become positive at the throat and display shallow wells rather than purely monotonic profiles. In contrast, in the Gaussian and Lorentzian cases $\Phi(x)$ always remained negative and monotonically increasing.  At the same time, the tangential pressures in the Chaplygin family follow the same overall pattern, i.e. negative near the throat and positive just outside, but their magnitude and localisation are strongly controlled by the nonlinearity parameter $\alpha$.  Increasing $\alpha$ both softens the redshift well and enhances the positive $p_t$ peak, showing that the Chaplygin-like EOS interpolates smoothly between the quasi–de Sitter limit and a more pronounced anisotropic regime while maintaining regularity and asymptotic flatness. In all cases considered, the NEC is violated only near the throat, where the Chaplygin nonlinearity is active, and tends to its de Sitter limit $\rho + p_r \to 0$ at large $x$.

\section{Conclusions and outlook}

We have constructed and analysed static, spherically symmetric wormhole geometries in a noncommutative geometry–inspired setting, where a minimal length $\sqrt{\theta}$ smears point sources into a Gaussian energy density. Starting from the regular shape function $b(r)$ associated with this profile, we derived model–independent relations that separate what the shape and the redshift functions control: (i) the sign of the NEC at the throat is fixed by $b'(r_0)$ according to \eqref{NEC}; (ii) for any regular redshift, a nonzero NEC–violating layer near the throat is unavoidable, and a sharp bound on $d\Phi/dr$ governs NEC restoration away from the throat as indicated by \eqref{eq:Phi-ineq}; and (iii) the width and depth of that layer are controlled by the near–throat gradient $\Phi_1$ via the slope formula \eqref{eq:NEC-slope} and the linear estimate for $r_c-r_0$ given by \eqref{eq:rc-deriv}, with the existence condition in \eqref{eq:rc-positivity}. 

Guided by these general results, we examined several representative redshift families (see Table~\ref{table:Phi}) chosen to sample distinct near–throat gradients and asymptotics while maintaining regularity, horizon avoidance, and asymptotic flatness. Across these families, we find a consistent trend, namely,  $8\pi(\rho+p_r)$ raises just outside the throat and the NEC–violating shell becomes progressively thinner, in agreement with equations \eqref{eq:NEC-slope}–\eqref{eq:rc-deriv}. We then recast this redshift engineering in matter terms by imposing a quasi–de Sitter EOS with localised Gaussian or Lorentzian perturbations, represented by the master relation \eqref{perturbation}. The resulting configurations realise minimally exotic wormholes, i.e. the NEC violation is strongly localised in magnitude near the throat, while $\rho+p_r\to 0^{-}$ asymptotically and the redshift function remains regular and asymptotically flat.

In addition, we explored a Chaplygin-like EOS that introduces a nonlinear coupling between pressure and density, inspired by the generalised Chaplygin gas. This extension produces redshift profiles that may become positive at the throat and develop shallow wells rather than purely monotonic curves, revealing the presence of local blueshift regions without forming horizons. The associated tangential pressures follow the same global pattern as in the Gaussian and Lorentzian cases—negative near the throat and positive just outside, but their magnitude and localisation are strongly controlled by the nonlinearity parameter $\alpha$. Increasing $\alpha$ simultaneously softens the redshift well and enhances the anisotropic peak, showing that the Chaplygin-like EOS provides a smooth interpolation between the quasi–de Sitter limit and a more pronounced anisotropic regime, all while preserving regularity and asymptotic flatness.

Beyond their mathematical interest, the configurations analysed in this work may have broader implications for the phenomenology of quantum gravity. The use of a noncommutative-geometry-inspired matter profile places these wormholes within a class of ultraviolet-regular spacetimes that arise naturally in various approaches to quantum gravity. 

Present bounds on the noncommutative length scale, $\sqrt{\theta} \lesssim 10^{-16}$ cm \cite{Nicolini2009IJMP}, imply that the throat radii $r_0 \sim \mathcal{O}(\sqrt{\theta})$ and the associated masses $M \sim \mu \sqrt{\theta}$ of our configurations are intrinsically microscopic and far below astrophysical scales. As a result, the wormholes constructed here are unlikely to leave direct imprints in current gravitational-wave or lensing observations. Instead, they should be viewed as controlled probes of the effect of a minimal length on wormhole energetics. Any experimental or observational bound on $\sqrt{\theta}$ obtained from quantum gravity–motivated tests, such as precision tabletop experiments that constrain minimal-length effects in quantum mechanics, or phenomenological studies of noncommutative-inspired black holes, translates directly into a bound on the allowed throat size and mass of these noncommutative wormholes. In particular, tighter upper limits on $\sqrt{\theta}$ push the wormhole configurations deeper into the microscopic regime, while any future positive indication of a finite minimal length would immediately fix the natural scale for the throat radius and mass in our solutions. At the cosmological level, a nonzero $\theta$ can modify early-universe physics through its role as an ultraviolet regulator. In that context, the geometries studied here provide explicit, ultraviolet-regular spacetimes that can serve as local models for nontrivial topological or wormhole-like fluctuations consistent with a minimal-length scale.

A natural next step is to extend this framework to slowly rotating wormholes within GR coupled to the same noncommutative–inspired matter. We aim to develop a controlled expansion in the angular–momentum parameter $J$, pushing the analytic treatment to $\mathcal{O}(J^3)$ (complemented by numerics as needed), thereby going beyond existing $\mathcal{O}(J^2)$ constructions \cite{Kashargin2008GC, Kashargin2008PRD}. Such an extension would clarify how rotation modifies the throat geometry, the near–throat NEC behaviour, and the global structure of these spacetimes. In parallel, stability diagnostics (radial perturbations and tidal constraints) and observational proxies (lensing or quasinormal signatures) can be evaluated within the same model–independent bounds established here.

\bibliography{QNMS}

\end{document}